\documentclass[12pt,a4paper]{article}

\usepackage{setspace} 
\usepackage[protrusion=true,expansion=true]{microtype}
\usepackage[utf8]{inputenc} 
\usepackage[T1]{fontenc} 
\usepackage[english]{babel} 
\usepackage{url} 
\usepackage{float}
\usepackage[adobe-utopia]{mathdesign} 
\usepackage[scaled=0.95]{helvet}

\usepackage{caption} 
\usepackage[EULERGREEK]{sansmath} 
\DeclareCaptionFont{sansmath}{\sansmath}
\DeclareCaptionLabelSeparator{pipe}{ | } 

\usepackage{amsmath}

\usepackage{commath}

\usepackage{jneurosci}

\usepackage{graphicx}

\usepackage{epstopdf}

\usepackage[squaren,thinspace,thinqspace]{SIunits}

\usepackage{xcolor}

\usepackage{soul}

\usepackage{setspace}

\usepackage[normalem]{ulem}

\usepackage{multicol} 

\usepackage{authblk}

\usepackage{enumerate}

\usepackage{tikz}  

\usepackage{hyperref} 
\hypersetup{colorlinks=true, citecolor=black!60, urlcolor=black, filecolor=black, linkcolor=black} 
\newcommand{\be}{\begin{equation}} 
\newcommand{\ee}{\end{equation}} 
\newcommand{\ben}{\begin{equation}} 
\newcommand{\een}{\end{equation}} 
\newcommand{\ba}{\begin{eqnarray*}}
\newcommand{\ea}{\end{eqnarray*}}

\newcommand{\by}{\begin{array}{cc}} 
\newcommand{\ey}{\end{array}} 
\newcommand{\bi}{\begin{itemize}} 
\newcommand{\ei}{\end{itemize}}

\def\slfrac#1#2{{\mathord{\mathchoice   
        {\kern.1em\raise.5ex\hbox{$\scriptstyle#1$}\kern-.1em
        /\kern-.15em\lower.25ex\hbox{$\scriptstyle#2$}}
        {\kern.1em\raise.5ex\hbox{$\scriptstyle#1$}\kern-.1em
        /\kern-.15em\lower.25ex\hbox{$\scriptstyle#2$}}
        {\kern.1em\raise.4ex\hbox{$\scriptscriptstyle#1$}\kern-.1em
        /\kern-.14em\lower.25ex\hbox{$\scriptscriptstyle#2$}}
        {\kern.1em\raise.2ex\hbox{$\scriptscriptstyle#1$}\kern-.1em
        /\kern-.1em\lower.25ex\hbox{$\scriptscriptstyle#2$}}}}}

\begin{document}

\newcommand\cb[1]{\textcolor{blue}{#1}} 
\newcommand\hk[1]{\textcolor{red}{#1}} 
\newcommand\jb[1]{\textcolor{magenta}{#1}} 

\newcommand\mmtext[1]{\textcolor{orange}{#1}} 
\newcommand{\mmrepl}[2]{\textcolor{orange}{\sout{#1} #2}} 

\newcommand\blue[1]{\textcolor{blue}{#1}} 
\newcommand\red[1]{\textcolor{red}{#1}} 
\newcommand\green[1]{\textcolor{green}{#1}} 
\newcommand\magenta[1]{\textcolor{magenta}{#1}}


\title{Broadly Heterogenenous Network Topology Begets Order-Based Representation by Privileged Neurons}

\author[1,2]{Christoph Bauermeister} \author[3]{Hanna Keren} \author[1,2]{Jochen Braun}

\affil[1]{Institute of Biology, Otto-von-Guericke University, Leipziger Str. 44 / Haus 91, 39120 Magdeburg, Germany} \affil[2]{Center for Behavioral Brain Sciences, Leipziger Str. 44, 39120 Magdeburg, Germany} \affil[3]{Network Biology Research Laboratory, Electrical Engineering, Technion - Israel Institute of Technology, Haifa 3200003, Israel}

\renewcommand\Authfont{\sffamily \bfseries \itshape \raggedright} \renewcommand\Affilfont{\sffamily  \mdseries \itshape\small \raggedright}

\date{} 

\maketitle 
\thispagestyle{empty} 
\let\oldthefootnote\thefootnote 
\renewcommand{\thefootnote}{\fnsymbol{footnote}} 
\footnotetext[1]{Correspondence: e-mail: \url{jochen.braun@ovgu.de}} \let\thefootnote\oldthefootnote

\begin{abstract}

How spiking activity reverberates through neuronal networks, how evoked and spontaneous activity interact and blend, and how the combined activities represent external stimulation are pivotal questions in neuroscience. We simulated minimal models of unstructured spiking networks {\it in silico}, asking whether and how gentle external stimulation might be subsequently reflected in spontaneous activity fluctuations. Consistent with earlier findings {\it in silico} and {\it in vitro}, we observe a privileged sub-population of `pioneer neurons' that, by their firing order, reliably encode previous external stimulation. We show that the distinctive role of pioneer neurons is owed to a combination of exceptional sensitivity to, and pronounced influence on, network activity. We further show that broadly heterogeneous connection topology -- a broad ``middle class'' in degree of connectedness -- not only increases the number of `pioneer neurons' in unstructured networks, but also renders the emergence of `pioneer neurons' more robust to changes in the excitatory-inhibitory balance. In conclusion, we offer a minimal model for the emergence and representational role of `pioneer neurons', as observed experimentally {\it in vitro}.  In addition, we show how broadly heterogeneous connectivity can enhance the representational capacity of unstructured networks.

\end{abstract}



\newpage

\section{Introduction} 

An important question in theoretical neuroscience is how externally evoked activity interacts with the spontaneous activity reverberating through neural networks.  Closely related questions are how network topology shapes this interaction (touching on structure-function relationships) and how the blending of evoked and spontaneous activity represents external stimulation (touching on the `neural code') \cite{spikes,decharms2000,thorpe,ponulak}. 

We address these issues by simulating {\it in silico} spiking neural networks (SNNs) with synaptic depression and different types of unstructured (random) connectivity.   Although representing a drastic oversimplification of cortical networks {\it in vivo}, randomly connected SNNs have contributed considerably to our understanding of neural function \cite{shepherd}.  They provide generic models for experimentally observed dynamics of cortical activity associated with phenomena such as short-term memory, attentional biasing, or decision-making  \cite{rolls1,rolls2}.  In addition, SNNs deepen our understanding of such phenomena because their activity dynamics can often be described analytically in terms of mean-field theory \cite{feng,gerstner}.

Randomly connected SNNs have long been studied experimentally by harvesting, dissociating, and culturing mature cortical neurons {\it in vitro} on the substrate of a multi-electrode array (MEA), which lets a  small fraction of spiking activity be monitored  ($O(0.1\%)$ of all neurons)\cite{morin,marom_shahaf_2002}.  The spontaneous activity of {\it in vitro} networks exhibits intrinsic fluctuations of all sizes, ranging from long quiescent spells to sudden synchronization events (`network spikes,' NS).  Depending on the balance of excitation and inhibition, {\it in vitro} networks may operate near, below, or above a regime of self-organized criticality (SOC), where fluctuations are distributed in a scale-free manner \cite{Bak,Jensen,Beggs,Pasquale,gigante}.  However, many experimental studies have elected to focus on a super-critical regime, in which periods of comparatively small activity fluctuations alternate with all-encompassing synchronization events. 
  
Several recent studies have investigated the gradual build-up of activity immediately prior to synchronization events (NS), as this build-up exhibits interesting features with possible functional implications.   Firstly, the growing activity propagates on reproducible paths, triggering a particular sequence of spikes in a certain subset of neurons (`pioneer neurons') \cite{eytan,Rolston2007}.  Secondly, the recruitment order of `pioneer neurons' is informative about prior perturbations by external stimulation \cite{shahaf}.  Specifically, when external stimulation is delivered to alternative sites, the stimulated site may be reliably decoded from the recruitment order of `pioneer neurons' \cite{kermany}.  Thirdly, the information encoded in the gradual build-up of activity may be propagated to other networks.  For example, the stimulated site in an upstream {\it in vitro} network, which sparsely projects to a downstream {\it in vitro} network, may be reliably decoded from the activity of the latter network \cite{Levy}.  Taken together, these observations raise the intriguing possibility that even unstructured neural networks express an order-based representation, encoding past external stimulation in the activity of a privileged class of  `pioneer neurons'.

Repeating `motifs' in the sequence of neuronal recruitment have been reported also {\it in vivo} in sensory cortex \cite{Luczak2007,Luczak2012a,Luczak2012b}, prefrontal and parietal cortex \cite{Peyrache2010,Rajan2016}, and in hippocampus \cite{Matsumoto2013,Stark2015}.  As the same `motifs' appear in spontaneous and evoked activity, they are considered an emergent property of local circuits \cite{Luczak2012b,Rajan2016}.  The possible functional significance of reproducible spike ordering, for example in the formation of memory patterns, is an active topic of research \cite{Contreras2013,Stark2015,Rajan2016}.

Numerous theoretical studies have investigated the collective spiking dynamics of unstructured networks \cite{tsodyks2,Loebel,Wiedemann,Persi2004a,Persi2004b,Vladimirski,Gritsun2008,Gritsun2010,Gritsun2011,Zbinden,Masquelier,Luccioli,gigante}.
Typically, these studies have combined  leaky integrate-and-fire neurons \cite{tuckwell} with frequency-dependent synapses \cite{tsodyks1}.  Different types of collective dynamics may be obtained (synchronous, asynchronous, critical, supra-critical, etc.), depending on balance of excitatory and inhibitory resources, as well as the dynamic characteristics of such resources (e.g., depletion rates)\cite{Poil}.   Several studies have focused on the supra-critical regime characterized by large synchronization events \cite{tsodyks2,Loebel,Wiedemann,Vladimirski,Luccioli,Masquelier,gigante}.

The emergence of `pioneer neurons' was first predicted by Tsodyks and colleagues \cite{tsodyks2}.   Ensuring heterogeneity by providing for different (effective) firing thresholds, these authors described a sub-population of neurons that fires reliably during the build-up towards a synchronization event.  Extending these results, `pioneers' with intermediate firing thresholds were shown to be critical for the generation of synchronization events \cite{Vladimirski}.  More detailed investigations suggested that `pioneers' tend to combine low firing thresholds with unusually dense out-going connectivity \cite{Zbinden}.  The formation of highly connected `leader' neurons could be favoured by activity-dependent plasticity \cite{Effenberger2015}.  

The primary focus of the present study lies on the interaction between  spontaneous activity dynamics and weak external stimulation.  Specifically, we investigated the conditions under which unstructured networks express `pioneer neurons' that form an order-based representation of prior stimulation.  An important contrast to previous studies is that we introduce neuronal heterogeneity by means of heterogeneous connection topology, rather than heterogeneous firing thresholds.

We show that the gradual build-up of activity towards a synchronization event proceeds in reproducible manner, recruiting identifiable `pioneer neurons' in a particular order, consistent with experimental findings \cite{eytan}.  We also show that this recruitment order is highly informative about the location of prior external stimulation, again consistent with experimental findings  \cite{shahaf,kermany}.  Additionally, we show that the neuronal heterogeneity required for `pioneer neurons' may arise from connection topology, specifically, by a broad ``middle class'' of neurons in terms of connectedness.   Finally, we show that broadly heterogeneous connectivity renders the collective dynamics more robust and ensures the presence of `pioneer neurons' over a wide range of excitatory-inhibitory architectures.

\section{Methods}


\subsection{Network design and parameters}

We simulated the collective activity of assemblies of $400$ excitatory and $100$ inhibitory neurons (leaky integrate-and-fire, LIF)  \cite{tuckwell}, connected randomly by means of conductance synapses with short-term dynamics \cite{tsodyks1}.  Spontaneous activity was evoked by a uniform and constant background current injected into all neurons.  As neuron models were identical, the only source of heterogeneity was the connectivity ($20 \,\, \%$ mean density).  Three types of connectivity were investigated: `homogeneous random' (Erd\"os-Rényi), `scale-free' \cite{Barabasi}, and `heterogeneous random'. 

A neural simulator was programmed in C and verified against existing simulators, as well as by reproducing the results of \cite{tsodyks2}.
Time was discretized in steps of $0.5 \,\, \mathrm{ms}$, except for power spectra, where steps of $0.1 \,\, \mathrm{ms}$ were used.  To ensure representative results, we investigated multiple realizations of every network architicture (typically more than 10).  Each type of connectivity expresses consistent behaviour, although event rates and average activity levels may vary from realization to realization.

\subsubsection{Neurons} The time-dependent membrane voltage $V$ was governed by the differential equation \be \frac{\mathrm{d} V}{\mathrm{d} t}(t)= \frac{E_L-V(t)}{\tau_m}+\frac{R_m I_b}{\tau_m}+\frac{R_m I_{\mathit{syn}}(t)}{\tau_m}, \label{Veq} \ee where $E_L$ is the leak reversal potential, $\tau_m$ is the membrane time constant, $R_m$ is  the membrane resistance, $I_b$ is the background current, and $I_{\mathit{syn}}$ is the synaptic  current, see below. Whenever the voltage reached the threshold $V_{\mathit{th}}$, it was reset immediately to $V_{\mathit{res}}$, where it remained for $\tau_{\mathit{ref}}$ (refractory  period).  The parameters of the neuron model were as follows: $E_L=-70 \,\, \mathrm{mV}$; $\tau_m =  30 \,\, \mathrm{ms}$ for excitatory neurons and $\tau_m = 10 \,\, \mathrm{ms}$ for  inhibitory neurons; $R_m=40 \,\, \mathrm{M \Omega}$ for excitatory neurons and $R_m=50 \,\, \mathrm{M \Omega}$ for inhibitory neurons; $I_b=525 \,\, \mathrm{pA}$ for excitatory neurons  and $I_b=420 \,\, \mathrm{pA}$ for inhibitory neurons; $V_{\mathit{th}}=-50 \,\, \mathrm{mV}$; $V_{\mathit{res}}=-65 \,\, \mathrm{mV}$; $\tau_{\mathit{ref}}=3 \,\, \mathrm{ms}$ for excitatory neurons and $\tau_{\mathit{ref}}=2 \,\, \mathrm{ms}$ for inhibitory neurons. Note that the background currents raise the equilibrium potential over the threshold level, ensuring spontaneous activity.   Note further that the model is defined without noise.  Initial membrane voltages were assigned randomly from the interval $[V_{\mathit{res}},V_{\mathit{th}}]$.  To avoid onset artefacts, the initial two seconds of activity were ignored.

\subsubsection{Synapses} The synaptic state is described by four time-dependent variables \cite{tsodyks1}: the instantaneous fractions of recovered, active, and inactive (synaptic) resources ($R(t)$, $E(t)$, and $I(t)$, respectively) and the fraction of resources $u(t)$ recruited by pre-synaptic spikes. These non-dimensional variables satisfy the following  equations: \ba \frac{\mathrm{d} R}{\mathrm{d}t}(t) & = & \frac{I(t)}{\tau_\mathit{rec}} - u(t+\epsilon)R(t-\epsilon) \rho(t), \\ \frac{\mathrm{d} E}{\mathrm{d}t}(t) & = & - \frac{E(t)}{\tau_I} + u(t+\epsilon)R(t-\epsilon) \rho(t), \\ \frac{\mathrm{d} I}{\mathrm{d}t}(t) & = & \frac{E(t)}{\tau_I} - \frac{I(t)}{\tau_\mathit{rec}}, \\ R(t)+E(t)+I(t) & = & 1, \\ \frac{\mathrm{d} u}{\mathrm{d}t}(t) & = & \frac{-u(t)}{\tau_{\mathit{facil}}} +U(1-u(t-\epsilon)) \rho(t), \ea where $\rho(t):= \sum\limits_{i} \delta(t-t_i)$ is the Dirac comb associated with the spike  train of the presynaptic neuron.  The axonal conduction delay was uniform and $0.5 \,\, \mathrm{ms}$.  $\tau_\mathit{rec}$ is  the recovery time constant; $\tau_I$ is the inactivation time constant; $\tau_{\mathit{facil}}$ is  the facilitation time constant; and $U$ is a parameter associated with resource utilization.   Parameter values are as follows (subscript 'ee' stands for 'excitatory-to- excitatory', 'ie' stands for 'excitatory-to-inhibitory', 'ei' stands for 'inhibitory-to- excitatory', 'ii' stands for 'inhibitory-to-inhibitory'): $\tau_{\mathit{I,ee}}=\tau_{\mathit{I,ie}}=3 \,\, \mathrm{ms}$ and $\tau_{\mathit{I,ei}}=\tau_{\mathit{I,ii}}=10 \,\, \mathrm{ms}$; the values for $U$, $\tau_\mathit{rec}$ and $\tau_{\mathit{facil}}$ were randomly chosen and hence varied from synapse to  synapse.   Values were chosen from Gaussian distributions with mean $U_\mathit{ee}=U_\mathit{ei}=0.3$; $U_\mathit{ie}=U_\mathit{ii}=0.04$; $\tau_\mathit{rec,ee}=\tau_\mathit{rec,ei}=0.8 \,\, \mathrm{s}$; $\tau_{\mathit{rec,ie}}=\tau_{\mathit{rec,ii}}=0.1 \,\, \mathrm{s}$; $\tau_{\mathit{facil,ie}}=\tau_{\mathit{facil,ii}}=1 \,\, \mathrm{s}$.   The standard deviation of each distribution was half the respective mean.  However, Gaussian distributions were clipped and restricted to a physically possible range ( i.e., positive values for time constants and values between zero and unity for $U$).  For $\mathit{ee}$- and $\mathit{ei}$-synapses, $\tau_{\mathit{facil}}$  was zero (no facilitation). 

The synaptic current $I_{\mathit{syn,i}}$ of the $i$-th neurons was \be I_{\mathit{syn,i}}(t)=g_{\mathit{exc,i}}(t)(E_{\mathit{exc}}-V(t))+g_{\mathit{inh,i}}(t)(E_{\mathit{inh}}-V(t)), \ee where the reversal potentials were chosen as $E_{\mathit{exc}}=0$ and $E_{\mathit{inh}}=-70 \,\, \mathrm{mV}$. The conductances $g_{\mathit{exc,i}}$ and $g_{\mathit{inh,i}}$ are given by \ba g_{\mathit{exc,i}}(t) & = & \sum\limits_{j \,\, \mathrm{exc}} w_{\mathit{ij}} E_{\mathit{ij}}(t) \\ g_{\mathit{inh,i}}(t) & = & \sum\limits_{j \,\, \mathrm{inh}} w_{\mathit{ij}} E_{\mathit{ij}}(t), \\ \ea where the sum is over all excitatory (inhibitory respectively) neurons. $w_{\mathit{ij}}$ is the matrix of synaptic weights, and $E_{\mathit{ij}}$ is the (time-dependent) matrix of resources in the active state. 

The assignment of neuron and synapse parameters was modelled on \cite{tsodyks2}.

\subsubsection{Connectivity matrix}   In homogeneous random (Erd\"os-Renyi) networks, each ordered neuron pair $(i,j)$ formed a synaptic connection $i\to j$ with $20 \,\, \%$ probability.  Over all neurons, the degree of connectivity thus followed a Gaussian distribution.  Scale-free networks were obtained with the `preferential attachment' procedure \cite{Barabasi}, such that connectivity followed a power-law distribution with a mean connectivity of $20 \,\, \%$.  Heterogeneous random networks were generated as follows.  Every neuron $i$ was individually assigned four random numbers,  $\lambda_{\mathit{pre,exc}}$, $\lambda_{\mathit{post,exc}}$, $\lambda_{\mathit{pre,inh}}$, and $\lambda_{\mathit{post,inh}}$, each drawn independently from the interval $[0,\delta]$, where $\delta=0.2$ is the mean connection  density.  In a second step, every ordered neuron pair $i,j$ was individually assigned two random numbers, $\xi$ and $\eta$, drawn independently from $[0,1]$.   An excitatory projection $j\to i$ was established, if neuron $j$ was excitatory and $\xi<\lambda_{\mathit{pre,exc}}$.  Similarly, an inhibitory projection $j\to i$ was established, if $j$ was inhibitory and $\xi<\lambda_{\mathit{pre,inh}}$.  Projections $i \to j$ were established if $j$ was excitatory and $\eta<\lambda_{\mathit{post,exc}}$, or if $j$ was inhibitory and $\eta<\lambda_{\mathit{post,inh}}$.  This procedure resulted in a random graph with mean connection density of $20 \,\, \%$.  Heterogeneity arises because  each neuron exhibits an individual connection density, with independent out-degree and in-degree.  

Established projections were assigned a synaptic weight $w_{\mathit{ij}}$, each chosen randomly and independently from a (clipped) Gaussian distribution with mean $\omega$ and  standard deviation $\omega/2$ (clipping ensured $w_{\mathit{ij}}>0$).  Not established projections were assigned $w_{\mathit{ij}}=0$.   Mean values were chosen such as to obtain spontaneous activity with pronounced synchronization events (`network spikes', see below) at rates of $O(10^0 \,\, \mathrm{Hz})$.   Specifically, we chose $\omega_\mathit{ee}=1150 \,\, \mathrm{pS}$, $\omega_\mathit{ei}=8500 \,\, \mathrm{pS}$, $\omega_\mathit{ie}=5 \,\, \mathrm{pS}$, $\omega_\mathit{ii}=200 \,\, \mathrm{pS}$ for homogeneous random networks; $\omega_\mathit{ee}=1450 \,\, \mathrm{pS}$, $\omega_\mathit{ei}=9500 \,\, \mathrm{pS}$, $\omega_\mathit{ie}=5 \,\, \mathrm{pS}$, $\omega_\mathit{ii}=200 \,\, \mathrm{pS}$ for scale-free networks; $\omega_\mathit{ee}=1000 \,\, \mathrm{pS}$, $\omega_\mathit{ei}=8500 \,\, \mathrm{pS}$, $\omega_\mathit{ie}=5 \,\, \mathrm{pS}$, $\omega_\mathit{ii}=200 \,\, \mathrm{pS}$ for heterogeneous random networks.

Almost all realizations of random connectivity resulted in spontaneous network activity including large synchronization events.  This was the case for $\approx 90 \,\, \%$ of the homogeneous random networks, $\approx 80 \,\, \%$ of the scale-free networks, and $\approx 100 \,\, \%$ of the heterogeneous random networks.   In the remaining realizations, spontaneous activity failed to ignite network spikes (in an all-or-nothing fashion).  The activity of our three networks was asynchronous irregular as shown by the large-frequency limit of the respective power spectral densities of the unfiltered population activity (cf. {\bf Figure~\ref{SpontAct} C}) \cite{spiridon}.

\subsection{Histograms and densities}

Histograms and densities were computed as follows. In {\bf Figure~\ref{SpontAct} B} and {\bf Figure~\ref{all-or-none} A} rectangular bins of, respectively, $100 \,\, \mathrm{ms}$ and $2$ were used. In all other cases, densities were estimated with Gaussian kernels.  Kernel width and sampling resolution for population activity (spike density) were $3 \,\, \mathrm{ms}$ and $0.1 \,\, \mathrm{ms}$, respectively.  For latency distributions, the corresponding values were $0.8 \,\, \mathrm{ms}$ and $0.1 \,\, \mathrm{ms}$, with approximately $400$ samples per kernel.  For voltage distributions they were $0.1 \,\, \mathrm{mV}$ and $10 \,\, \mathrm{\mu V}$, with $20,000$ samples.   For the fraction of recovered resources $R$, kernel width and voltage distribution were $0.05$ and $0.01$, with $O(10^4)$ samples.

Recovered resources $R_i(t)$ of neuron $i$ were averaged over synapses $i\to j$ to $N_i$ post-synaptic neurons $j$
\be
R_i(t)=\frac{\sum_j R_{ji}(t)}{N_i},
\ee
where $R_{ji}(t)$ is recovered resources of synapse $i\to j$ at time $t$.  Densities $p_i(R)$ were established over all time points {\it excluding} NS ({\it i.e.}, time-points more than $35 \,\, \mathrm{ms}$ before or after a NS).

Population activity is understood to be activity of the excitatory subpopulation and is sometimes given as {\it absolute} activity (in $\mathrm{Hz}$) and sometimes as {\it relative} activity ({\it i.e.}, in units of the average activity level).

 \subsection{Network spikes, peak activities, and synchrony thresholds}
 
Network spikes (NSs) are large bursts of excitatory activity separated by long periods of low activity.  We defined NS with respect to a high threshold (half the maximal activity): $\theta_\mathrm{high} = 0.5 \cdot \mathrm{max}[A(t)]$,  where $A(t)$ is excitatory activity in $\mathrm{Hz}$.   Beginning and end $[t_i,t_f]$ of a NS were defined as successive crossings of $\theta_\mathrm{high}$ by $A(t)$ from below and from above.  For each NS, we determined duration $t_{\mathit{NS}}=t_f-t_i$ and peak activity $A_{\mathit{max}}$. 

Note that this definition captures only large bursts of activity.   Smaller fluctuations were detected with a lower threshold $\theta_\mathrm{low} = 1.1 \cdot \mathrm{mean}(A)$.  The distribution of peak activities $A_{\mathit{max}}$ could be monomodal or bimodal.  Bimodal distributions indicated `all-or-none' synchronization events, consistent with a supercritical regime \cite{gigante}.  Typically, probability density was divided between low values ($< 10$ times mean activity) and high values ($>40$ times mean activity), with zero density in between.

The largest `low' values constitute a lower bound for the `threshold' of NS initiation, as any intermediate values must have resulted in `runaway' amplification to a `high' value.  Specifically, we determined this `threshold' as the largest observed value of the lower concentration of probability mass. In sufficiently long simulations, this empirical value should provide a close lower bound for the `true' threshold.

To characterize activity between NS, we exclude NS by omitting $35 \,\, \mathrm{ms}$ of activity before and after peak activity.  This value reflects the shape of NS, which is highly stereotypical (not shown).

\subsection{Encoding of external stimulation}

To assess the extent to which network activity encodes external stimulation, we perturbed spontaneous network activity in some simulations.  External stimulation targeted particular subsets of excitatory neurons (10 or 30 randomly selected neurons) and forced a single spike in each target neuron.  Each subset of targets was considered a `stimulation site'  and up to 12 non-overlapping sites were used.   Subsequent network activity ({\it i.e.}, $100 \,\, \mathrm{ms}$ of activity, exclusive of the forced spikes) was characterized in terms of four features, following \cite{kermany}. Two features were based on firing rates $a_i(t)$ of neurons $i$: temporal profile of population activity $A(t) = \sum_i a_i(t)$, and spatial profile of population activity $A_i = \int a_i(t) dt$.  Two further features were based on spiking activity of neurons $i$ before the subsequent NS: timing of first spikes $t_i$, and rank order of first spikes $o_i$.  Rank order was obtained by sorting negative spike latencies with respect to the subsequent NS (for example, the negative latency vector $(-20 \,\, \mathrm{ms},-10 \,\, \mathrm{ms}, -15 \,\, \mathrm{ms}, -17 \,\, \mathrm{ms})$ would yield the rank order vector $(1,4,3,2)$). 

To analyze the information encoded by different activity features, simulates were divided into a training and a test set.  Following \citetext{kermany}, the training set was used to train a support vector machine (SVM, linear kernel) to classify the stimulated location on the basis a particular feature.  The test set was used to determine classification performance (fraction of correct classifications of the stimulated site), providing a lower bound for the `true' information about stimulation site encoded by a particular activity feature.

 \subsection{Identification of pioneer neurons} 
 \label{semi-analytic}
 
Pioneer neurons fire reliably in advance of NSs.  In long simulations, pioneers could be identified as excitatory neurons which establish a characteristic individual spike density relative to peak NS activity.  Unfortunately, this was not practical for comparing thousands of networks with different connectivities.  To overcome this difficulty, we developed a computationally more efficient semi-analytic procedure, which produced comparable results. This semi-analytic procedure was validated by comparison with direct measurements of latency distributions (by long simulations).
The acceleration factor was quite high ($O(10^3)$), because only quantities accessible from short simulations (few seconds) need to be determined empirically. In contrast, directly measuring first-spikes densities requires much longer simulations, because NSs appear on a time scale of $1 \,\, \mathrm{s}$, and because many NSs need to be observed for determining spike latency distributions directly.

The first-step of the semi-analytic approach was to model NS initiation as an exponential rise activity.  Individual neuron activity was  assumed to rise from its measured baseline level to a common target level.   Specifically, firing of neuron $i$ was modeled as an inhomogeneous Poisson process with time-dependent rate $\alpha_i(t)$ (`activation function').  Given measured baseline activity $r_{\mathit{0,i}}$ and target activation $A$, the activation function of excitatory neurons was given by: \be \alpha_{i}(t) = r_{\mathit{0,i}} + (A-r_{\mathit{0,j}}) \exp{\left( - \frac{|t|}{\tau} \right)}. \ee  Inhibitory neurons did not participate the exponential rise: $\alpha_{i}(t) = r_{\mathit{0,i}}$.   Target activation $A$ was ten times average activation and time-constant $\tau$ was $0.01 \cdot \left\langle \mathit{INSI} \right\rangle$, where $\left\langle \mathit{INSI} \right\rangle$ is the measured average inter-network-spike interval.

Given exponentially rising activity of presynaptic neurons $j$ and dynamical synaptic connectivity, we sought to obtain analytically the expected firing density of postsynaptic neurons $i$ during this period.  To this end, we need to solve a bespoke time-dependent Fokker-Planck equation (FPE) for each excitatory neuron, taking into account the presynaptic firing densities of all incoming projections. To this end we determined, firstly, the expected cumulative time-course of inhibitory and excitatory synaptic conductances in postsynaptic excitatory neurons $i$ and, secondly, the time-dependent distribution of membrane potentials $p_i(t,V)$ of such neurons $i$. From $p_i(t,V)$ it is then only one more step to the latency distribution of neuron $i$ (see further below).

Synaptic conductances follow from $\left\langle R_{\mathit{ij}}(t) \right\rangle$ and $\left\langle E_{\mathit{ij}}(t) \right\rangle$.  Recalling that synapses $j \rightarrow i$ onto excitatory neurons $i$ do not show faciliation, we obtain $\left\langle R_{\mathit{ij}}(t) \right\rangle$ from the ODE \be \frac{\mathrm{d} \left\langle R_{\mathit{ij}} \right\rangle}{\mathrm{d} t}(t)= \frac{1-\left\langle R_{\mathit{ij}} \right\rangle(t)}{\tau_{\mathit{rec,ij}}}-U_{\mathit{ij}} \left\langle R_{\mathit{ij}} \right\rangle(t) \alpha_j(t), \label{eq1} \ee where the inactivation time-scale has bee neglected. The analytical solution is: \ba \left\langle R_{\mathit{ij}} \right\rangle (t) & = & 1-(1-C_{\mathit{ij}}) \exp{\left(-\frac{(t-t_0)}{\tau_{\mathit{rec,ij}}}- U_{\mathit{ij}} \int_{t_0}^t \mathrm{d} t' \alpha_j(t')\right)} \\ & - &  U_{\mathit{ij}} \int_{t_0}^t \mathrm{d}t' \alpha_j(t') \exp{\left( -\frac{(t-t')}{\tau_{\mathit{rec,ij}}} - U_{\mathit{ij}} \int_{t'}^{t} \mathrm{d} t'' \alpha_j(t'') \right)}, \ea where $C_{\mathit{ij}}=\left\langle R_{\mathit{ij}}(t_0) \right\rangle$ is the initial condition at $t_0=-0.1 \,\,\mathrm{s}$ (which is more negative than $-\tau$ in virtually all cases).  Accordingly, the initial condition of $\left\langle R_{\mathit{ij}} \right\rangle$ was chosen as the stationary solution of equation (\ref{eq1}):\be C_{\mathit{ij}}=\frac{1}{1+U_{\mathit{ij}} r_{\mathit{0,j}} \tau_{\mathit{rec,ij}}}.\ee  From $\left\langle R_{\mathit{ij}}(t) \right\rangle$, we computed $\left\langle E_{\mathit{ij}}(t) \right\rangle$ as the solution of \be \frac{\mathrm{d} \left\langle E_{\mathit{ij}} \right\rangle }{\mathrm{d}t}(t)  =  - \frac{\left\langle E_{\mathit{ij}} \right\rangle (t)}{\tau_I} + U_{\mathit{ij}} \left\langle R_{\mathit{ij}} \right\rangle (t) \alpha_j(t), \ee which is \ba \left\langle E_{\mathit{ij}} \right\rangle (t) & = &  D_{\mathit{ij}} \exp{\left( - \frac{t-t_0}{\tau_{\mathit{I,ij}}}  \right) } \\ & + & U_{\mathit{ij}} \int\limits_{t_0}^t \mathrm{d}t' \left\langle R_{\mathit{ij}} \right\rangle (t') \alpha_j(t') \exp{\left( - \frac{t-t'}{\tau_{\mathit{I,ij}}} \right) }, \ea where $D_{\mathit{ij}}$ is the initial condition \be D_{\mathit{ij}}=\frac{\tau_{\mathit{I,ij}} U_{\mathit{ij}} r_{\mathit{0,j}}}{1+U_{\mathit{ij}} r_{\mathit{0,j}} \tau_{\mathit{rec,ij}}}. \ee From $\left\langle E_{\mathit{ij}}(t) \right\rangle$, we obtained cumulative synaptic conductances in excitatory neurons $i$ as \be \left\langle g_{\mathit{exc,i}} \right\rangle (t) = \sum\limits_{j \,\, \mathit{exc}} w_{\mathit{ij}} \left\langle E_{\mathit{ij}} \right\rangle (t), \ee and \be \left\langle g_{\mathit{inh,i}} \right\rangle (t) = \sum\limits_{j \,\, \mathit{inh}} w_{\mathit{ij}} \left\langle E_{\mathit{ij}} \right\rangle (t). \ee 

Given the time-course of synaptic conductances, the expected firing density follows from a Fokker-Planck equation (FPE) for the membrane potential (in the `diffusion limit', \cite{rolls2}). In order to `derive' this FPE, we first observe that we may ignore the firing threshold and subsequent reset and compute the distribution of a hypothetical `star-voltage', which may be non-zero even above threshold.  There are two reasons why this hypothetical star-voltage is sufficient for our purposes.  Firstly, when neurons do not fire between NSs (as pioneers do), the star-voltage is identical to the actual voltage.  Secondly, when neurons do fire between NSs, we are interested only in the \it first \rm spike, so that the subsequent evolution is irrelevant.  The next step in the derivation of the FPE is to obtain from equation (\ref{Veq}) the \it stationary \rm membrane star-voltage. This reads \be \left\langle V_i \right\rangle_{\mathit{ss}} = \frac{E_L+R_{\mathit{m,i}} I_{\mathit{b,i}}+R_{\mathit{m,i}} \left( E_{\mathit{exc}} \left\langle g_{\mathit{exc,i}} \right\rangle_{\mathit{ss}}+E_{\mathit{inh}} \left\langle g_{\mathit{inh,i}} \right\rangle_{\mathit{ss}} \right) }{1+R_{\mathit{m,i}} \left( \left\langle g_{\mathit{exc,i}} \right\rangle_{\mathit{ss}}+\left\langle g_{\mathit{inh,i}} \right\rangle_{\mathit{ss}} \right) }, \label{Vstat} \ee where we assumed statistical independence between conductance and voltage. We then arrive at the following phenomenological stochastic differential equation for the star-voltage of neuron $i$: \be \frac{\mathrm{d} V_i }{\mathrm{d} t}(t) = \frac{\mu_i(t)-V_i(t) }{\tau_{\mathit{m,i}}} + \sqrt{\frac{2D}{\tau_{\mathit{m,i}}^2}} \xi(t), \label{wn} \ee where the drift term $\mu_i(t)$ is obtained in analogy to equation (\ref{Vstat}) as \be \mu_i(t) = \frac{E_L+R_{\mathit{m,i}} I_{\mathit{b,i}}+R_{\mathit{m,i}} \left( E_{\mathit{exc}} \left\langle g_{\mathit{exc,i}} \right\rangle (t) +E_{\mathit{inh}} \left\langle g_{\mathit{inh,i}} \right\rangle (t) \right) }{1+R_{\mathit{m,i}} \left( \left\langle g_{\mathit{exc,i}} \right\rangle (t) +\left\langle g_{\mathit{inh,i}} \right\rangle (t) \right) }, \ee and where $\xi(t)$ denotes Gaussian white noise. Equation (\ref{wn}) has the correct stationary solution (\ref{Vstat}) under steady-state conditions, and by choosing the noise strength $D$ in such a way that the stationary standard deviation $\sigma_i$ of $V_i$ is identical to the empirically determined one, one obtains an equation which describes rather well the star-voltage dynamics of neuron $i$, despite the fact that $D$ is assumed constant during the recruitment process.

The time-dependent FPE corresponding to (\ref{wn}) \be \frac{\partial p_i}{\partial t}(t,V) = \frac{1}{\tau_{\mathit{m,i}}} \frac{\partial (V-\mu_i(t))p}{\partial V}(t,V) + \frac{D}{\tau_{\mathit{m,i}}^2} \frac{\partial^2 p}{\partial V^2}(t,V) \ee needs to be supplemented by the boundary condition $p_i(t,V_{\mathit{th}})=0$ \cite{rolls2}.  As we are only interested in \it first \rm spikes, the usual reinsertion flux at the reset voltage was omitted.  We also neglected the boundary condition at threshold. The solution to the above FPE then reads \be p_i(t,V)=\frac{1}{\sqrt{2 \pi \sigma_i^2}} \exp{\left[ - \frac{\left( V_i- \mu_i(t_0)  \exp{\left( - \frac{(t-t_0)}{\tau_{\mathit{m,i}}} \right)} - \frac{1}{\tau_{\mathit{m,i}}} \int_{t_0}^{t} \mathrm{d} t' \mu_i(t') \exp{\left( - \frac{(t-t')}{\tau_{\mathit{m,i}}}\right)} \right)^2 }{2 \sigma_i^2} \right]}. \ee We then obtained the expected density of first spikes as $f_i(t)$ as \be f_i(t)=\mathrm{max} \left( 0 , -  \frac{\mathrm{d}}{\mathrm{d} t} \left( \int_{-\infty}^{V_\mathrm{th}} p_i(t,V) \mathrm{d} V \right) \right). \ee
because the spike-density equals the probability flux through threshold, which is give by the above expression. The half-wave rectification suppresses negative spike-densities, which may occur during the the falling phase of the NS due to the neglected boundary condition.

This semi-analytical procedure was validated by comparing the semi-analytically and numerically obtained densities for first-spike latencies.

\subsection{Silencing of neurons}

To assess the relative importance of different subsets of excitatory neurons, we wished to render ineffective the members of any particular subset.  To do so, we retained the spikes of such neurons but suppressed all post-synaptic effects.  A reduced frequency of NS in partially de-afferentiated networks revealed the relative importance of the manipulated subset of neurons.

\subsection{Spike-triggered average population activity}

To assess the relation between population activity and individual neuron spikes, we computed the `spike-triggered deviation' $\Gamma_i(\tau)$ as follows
\be
\Gamma_i(\tau) = \frac{\int \left( A(t) - \left\langle A \right\rangle \right) \rho_i(t-\tau) \mathrm{d} t}{\int \rho_i(t) \mathrm{d}t}
\ee
where $A(t)$ is time-dependent population activity, $\left\langle A \right\rangle$ is its temporal mean, $\rho_i(t)$ is the spike sequence (Dirac comb) of neuron $i$, and $\tau$ is the latency between activity time $t$ and spike time $t-\tau$.  The computation was restricted to periods in between NS and the normalization term $\int \rho_i(t) \mathrm{dt}$ is the number of spikes fired by neuron $i$ between NS.  In principle, $\Gamma_i(\tau)$ measures {\it influence} on ($\tau>0$), as well as sensitivity to ($\tau<0$), population activity of neuron $i$.  However, as many neurons spike only shortly before NS, we mostly obtain information about negative latencies, that is, about sensitivity to population activity.  For this reason, the spike-triggered deviation in  {\bf Figure~\ref{first_hist2} C} is restricted to negative latencies.  Moreover, it is defined only for (sorted) neuron ID $>260$, as less active neurons never spike between NS.

\subsection{Estimation of post-synaptic effects}

To estimate the differential effect of neuron $i$ on post-synaptic neurons $j$ throughout the network, we proceeded as follows. For every synaptic target $j$, we formed the difference $W_\mathit{ji}\equiv V'_j - V_j$ between the hypothetical star-voltage $V'_j(t)$ that would have resulted from a single additional spike of neuron $i$ at time $t_\mathrm{sp}$ and the actual the star-voltage $V_j(t)$, which (neglecting conduction delays) may be approximated as
\be
W_\mathit{ji}(t) \approx \frac{\tau_I U w R_m (E_\mathrm{exc}-V_\mathrm{th})}{(\tau_m - \tau_I) (1 + U \tau_\mathrm{rec} \nu_i)} \Theta(t-t_\mathrm{sp}) \left[  e^{ -(t-t_\mathrm{sp}) / \tau_m } - e^{ -(t-t_\mathrm{sp})/\tau_I }\right]. \label{W}
\ee
where $\nu_i$ is the asynchronous firing rate of neuron $i$ (between NS) and $\tau_I$, $U$, $w$, $\tau_\mathrm{rec}$ are parameters of the synapse in question.  The expression for $W_\mathit{ji}(t)$ peaks at time
\be
t_\mathrm{max} = \frac{\tau_m \tau_I }{(\tau_m - \tau_I)} \log{ \left( \frac{\tau_m}{\tau_I} \right)}.
, \label{tmax}
\ee
so that the post-synaptic potential in neuron $j$ that is triggered by the additional spike in neuron $i$ at time $t_\mathrm{sp}$ is $W_\mathit{ji}(t_\mathrm{max})$.  The cumulative post-synaptic effect of all spikes in neuron $i$ is given by the stationary limit 
\be
\left\langle W_\mathit{ji}\right\rangle_\mathrm{ss} = \frac{R_m w (E_\mathrm{exc}-V_\mathrm{th}) \tau_I U \nu_i}{1+U \tau_\mathrm{rec} \nu_i}.
\ee
which is approximately equal to $\tau_m \cdot \nu_i \cdot W_\mathit{ji}(t_\mathrm{max})$.

In {\bf Figure~\ref{depression} BC}, the differential effects of neuron $i$ are averaged over all $N_i$ post-synpatic neurons $j$
\begin{equation*}
\mathrm{PSP}_i \equiv \frac{1}{N_i} \, \sum_j W_{ji}(t_\mathrm{max}), \qquad\qquad \left\langle \mathrm{PSP}_i \right\rangle_\mathrm{ss} \equiv \frac{1}{N_i} \, \sum_j \left\langle W_{ji} \right\rangle_\mathrm{ss}.
\end{equation*}

\subsection{Modification of Levenshtein edit distance}

To quantify dissimilarity in the rank order or `first spikes' observed in different contexts, we modified the Levenshtein edit distance \cite{levenshtein} used in previous studies \cite{shahaf}.  Whereas the Levenshtein metric is useful for strings with same and/or different `letters', in the present situation all rank order strings contain the same `letters' (because all neurons fire at least one spike and rare missing spikes can be `filled in' at the highest rank).   Now consider two strings $s_1s_2 \dots s_n$ and $s_{\pi(1)} s_{\pi(2)} \dots s_{\pi(n)}$, where $\pi$ is an appropriately chosen permutation. Then number of inversions $L$, which is the number pairs $(i,j)$ such that $i<j$ but $\pi(i)>\pi(j)$, ranges from  $0$ (if strings are identical) to $L=\frac{n(n-1)}{2}$ (if strings are inverted).  Accordingly, we adopted
\be
L_n = \left(  1 - \frac{2L}{n(n-1)} \right) \, 100 \%
\ee
as normalized measure of similarity.

\section{Results}

We begin by describing macroscopic network dynamics, focusing on spontaneous fluctuations of activity and on the effect of gentle external stimulation (\ref{macroscopic_behaviour} Macroscopic behaviour).   Next we characterize `pioneer neurons' in terms of their sensitivity to, and influence on, activity fluctuations and in terms of their contribution to network amplification (\ref{pioneer_neurons} Pioneer neurons).  We then  compare the representation of gentle external stimulation by different aspects of population activity, including by pioneer neurons (\ref{order_based} Order-based representation).  Lastly, we show that some types of random connectivity express the macroscopic and microscopic activity regimes in question more robustly and reliably than other types (\ref{broadly_heterogeneous} Broadly heterogeneous connectivity).

\subsection{Macroscopic behaviour}
\label{macroscopic_behaviour}

This section describes spontaneous activity and activity evoked by gentle external stimulation.  Following previous studies \cite{gigante,tsodyks2,Masquelier,Loebel,Vladimirski,Wiedemann,Luccioli}, we focused on network architectures that combine low average activity with bimodal activity fluctuations (`all-or-nothing' synchronization events).

\subsubsection{Spontaneous activity} 
\begin{figure}[!ht]   \centering   \includegraphics[]{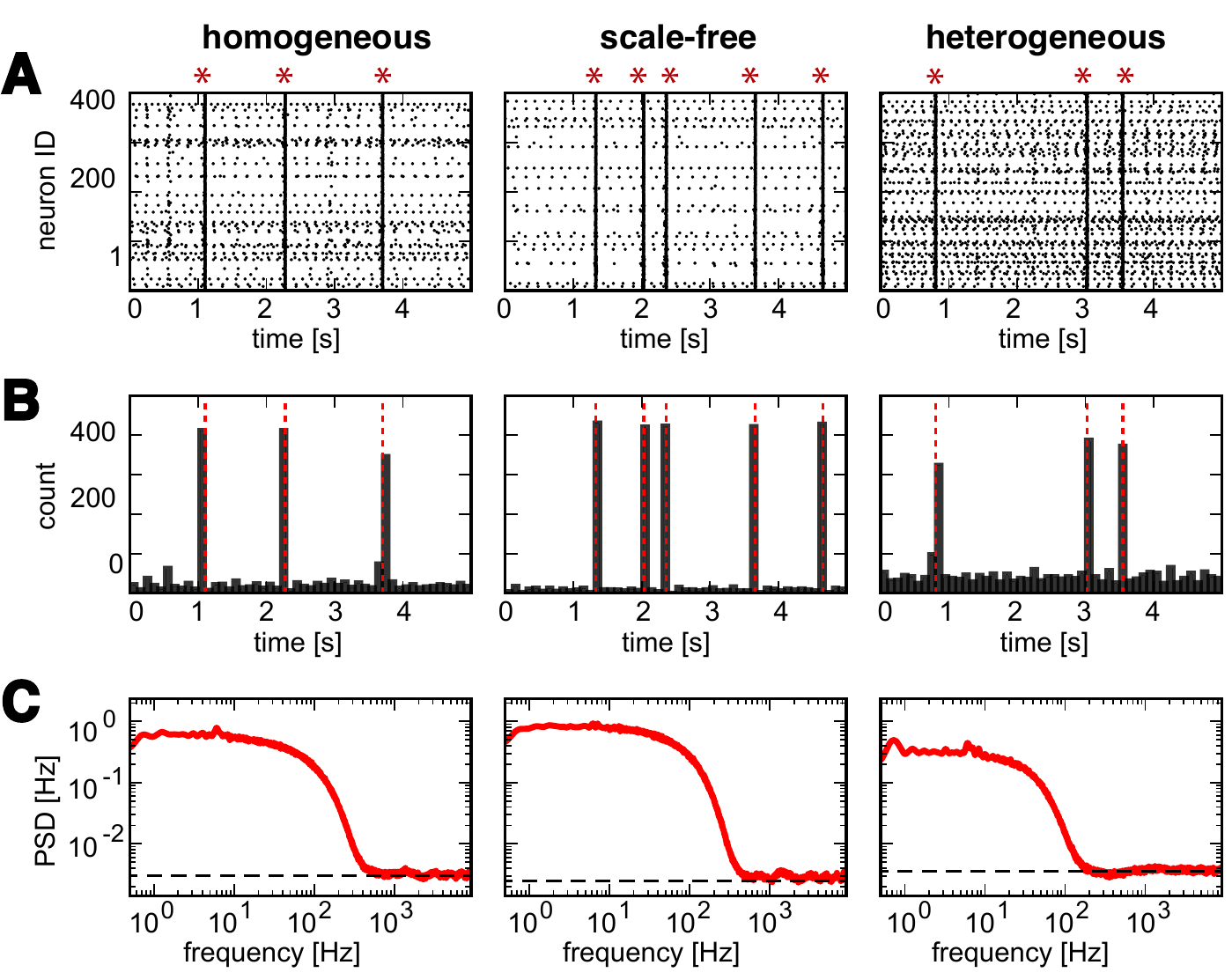}   \caption{$\mathbf{A}$: Spike rasters of excitatory neurons for the three network types (representative examples) with NS (red stars).  $\mathbf{B}$: Corresponding spike counts (bin width $100 \,\, \mathrm{ms}$) with NS (dashed red lines). $\mathbf{C}$: Power spectral densities of unfiltered excitatory population activity, with high-frequency limit $\frac{\nu}{N}$ for asynchronous irregular spiking (dashed black lines), where $\nu$ is activity in $\mathrm{Hz}$ and $N$ is neuron number.} \label{SpontAct} \end{figure}

Representive periods of spontaneous activity are illustrated in {\bf Figure~\ref{SpontAct} A and B}.  The generally low level of activity is briefly interrupted by spontaneous synchronization events (network spikes, NS), which recruit nearly all excitatory neurons at least once.   Network spikes occur at irregular intervals (coefficient of variation $c_v\approx 0.6$) and with frequencies $O(1 \,\, \mathrm{Hz})$ (due to a suitable balance of excitation-inhibition, see Methods).   Although the network is deterministic, finite-size noise ensures asynchronous irregular spiking, as shown by the power spectral density of spiking activity  {\bf Figure~\ref{SpontAct} C}, which approaches the expected theoretical value for large frequencies \cite{spiridon}.  

A telling characteristic of spontaneous population activity is the size distribution of positive fluctuations {\bf Figure~\ref{all-or-none} A}.     Following earlier studies, we investigated networks with bimodally distributed fluctuations, where larger fluctuations constituted NS and distinctly smaller fluctuations represented the intervening periods.  This bimodal dynamics allowed NS to be identified unambiguously.   In heterogeneous random networks, larger and smaller fluctuations were separated less widely than in other network types.   Rates of excitatory neuron spikes and NS were comparable ($2.1 \pm 3.0 \,\, \mathrm{Hz}$ and $1.7 \pm 3.1 \,\, \mathrm{Hz}$, respectively).  In homogeneous random and scale-free networks, NS were considerably larger, with many neurons contributing multiple spikes to each NS.  Accordingly, the rate of neuron spikes ($1.0 \pm 1.5 \,\, \mathrm{Hz}$ and $1.2 \pm 1.1 \,\, \mathrm{Hz}$, respectively) was several times larger than the rate of NS ($0.4 \pm 1.6 \,\, \mathrm{Hz}$ and $0.2 \pm 1.1 \,\, \mathrm{Hz}$). In all three network types, inhibitory neurons fired continuously at $30 \pm 1 \,\, \mathrm{Hz}$.

\subsubsection{Evoked activity} 

\begin{figure}[!ht]   \centering   \includegraphics[]{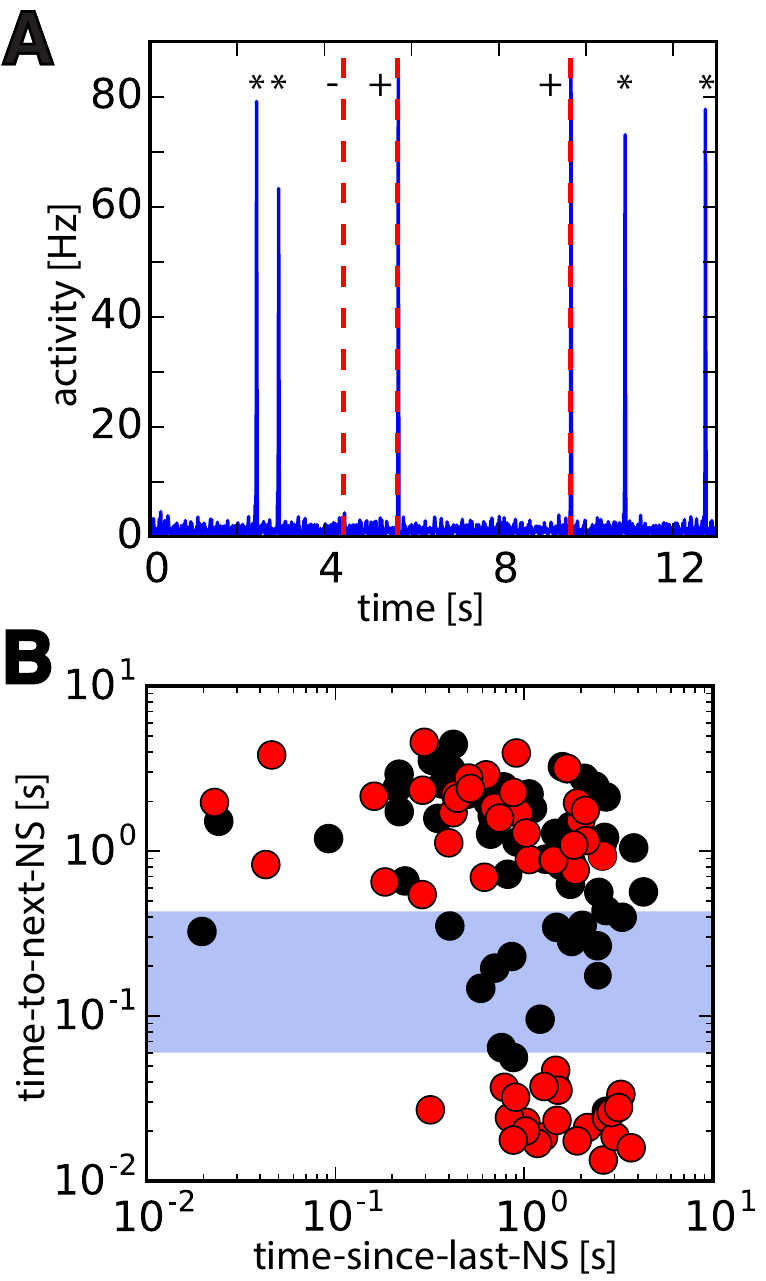}   \caption{ $\mathbf{A}$: Superposition of spontaneous and evoked activity (example sequence).  External stimulation forced simultaneous spikes in $5$ randomly chosen excitatory neurons, at random time-points (Poisson rate $0.4 \,\, \mathrm{Hz}$), here marked by dashed red lines.  Stimulation that succeeded (failed) to evoke a NS is marked by `$+$' (`$-$').  Spontaneous NS are denoted by '$\star$'. $\mathbf{B}$: Classification of evoked and spontaneous NS.  For each stimulation event, time-to-next-NS is plotted against time-since-last-NS (red dots).   For comparison, a null distribution is shown for an identical number of randomly timed, surrogate events (black dots).   Stimulation events (red dots) form two distinct clusters (above and below blue bar), permitting us to classify stimulation events with high probability as either successful (below) or unsuccessful (above).  In contrast, surrogate events are distributed continuously. Based on $120\,\,\mathrm{s}$ simulation of a heterogeneous random network, with stimulation rate equal to spontaneous NS rate.} \label{stimulation} \end{figure}

A representative combination of spontanous and evoked activity is shown in {\bf Figure~\ref{stimulation} A}.   External stimulation was delivered by forcing \it simultaneous \rm spikes in a small number of excitatory neurons.   As the effectiveness of stimulation  varied considerably with the number and identity of target neurons, we simulated multiple ($O(10^1)$) network realizations to ensure representative results.  To keep evoked and spontaneous NS as comparable as possible, we opted for a comparatively `gentle' stimulation targeting a small set of neurons ($O(10^1)$).

To classify NS as either evoked or spontaneous, we adapted the method of \citetext{Poster}, which relates time of stimulation to the timing of both previous and next NS.  Specifically, plotting the interval to the next network spike (`time-to-next-NS'), against the interval since the previous network spike (`time-since-last-NS'), one observes a significant negative correlation ({\bf Figure~\ref{stimulation} B}, red dots).  The reason is that evoked NS are more likely {\it long after} the previous NS (due to the refractoriness of excitatory synaptic resources) and {\it shortly after} stimulation.  Neither dependence is expected to hold for randomly timed `surrogate events' (black dots).  Interestingly, the observed distribution of stimulation events (red dots) formed two distinct clusters, which was not the case for surrogate events (black dots).  This difference, which was obtained consistently in all simulations, reveals a suppressive effect of stimulation on spontaneous NS, within a certain range of latencies.  Presumably, even unsuccessful stimulation consumes some synaptic resources, thus increasing time-to-next-NS.  This bimodal clustering of stimulation events helps classification.  We classify as `unsuccessful' stimulation events with long time-to-next-NS (above blue bar) and as `successful' stimulation events with short time-to-next-NS (below blue bar).


 \clearpage

\subsection{Pioneer neurons} 
\label{pioneer_neurons} 

We now turn to microscopic activity that differentiates individual excitatory neurons, especially `pioneer neurons'.   In our simulated networks,  the only possible source of differentiation was random variability in connectivity.   In general, this microscopic differentiation was qualitatively similar for all types of connectivity.   However, as detailed in the section `Broadly heterogeneous connectivity', the microscopic phenomenology is considerably more robust and reproducible in heterogeneous random networks, than in homogeneous random or scale-free networks.  The latter two typically require fine-tuning of parameters.  For this reason, we document the microscopic behavior for heterogeneous random networks.  In particular, all plots in section $3.2$ and $3.3$ were generated with heterogeneous connectivity.

\begin{figure}[!ht]   \centering   \includegraphics[]{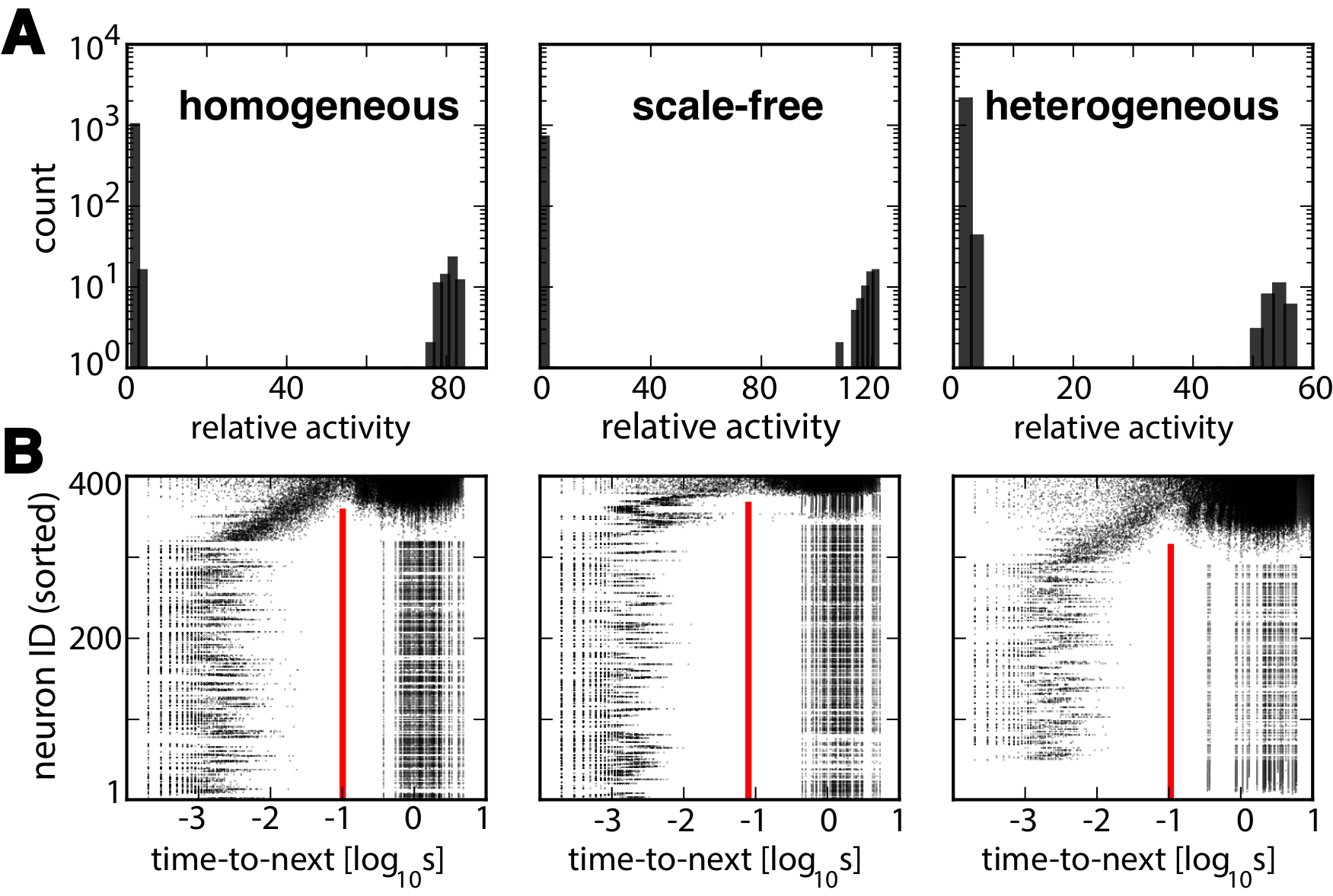}   \caption{$\mathbf{A}$: Histogram of peak-activation values during spontaneous activity fluctuations ($100\,\,\mathrm{s}$ simulations).  Relative activity in multiples of mean firing rate $A_\mathrm{mean}$ (from left: $1.20 \,\, \mathrm{Hz}$, $0.99 \,\, \mathrm{Hz}$, and $1.43 \,\, \mathrm{Hz}$, respectively). $\mathbf{B}$: Individual neuron spikes, relative to next NS, during $100\,\,\mathrm{s}$ of spontaneous activity.   Excitatory neurons are sorted vertically by mean firing rate (sorted neuron ID), with the least active neuron at bottom and the most active neuron on top.   Individual neuron spikes are represented by black dots.  For most neurons, spikes fall into two distinct classes: shortly before or long before the next NS (left and right columns of black dots, respectively).  A heuristic latency criterion (red lines) readily distinguishes these classes.  Thus, the `first spike during a NS'  is well defined ({\it i.e.}, rightmost dots in left column), for all but the most active neurons.} \label{all-or-none} \end{figure}

Due to random variations of connectivity, excitatory neurons fire with consistently different mean rates.  The least active neurons never discharge, many neurons fire one spike per NS, and the most active neurons fire several spikes per NS.   In most neurons, nearly all spikes are associated with NS and very few spikes occur during intervening periods.   Due to this close association with NS, one may identify the `first spike' of a particular neuron within a particular NS, at least for all but the most active neurons.  This is illustrated in {\bf Figure~\ref{all-or-none} B}, which shows a raster of individual neuron spikes, relative to peak activity of the next NS.   Individual neuron spikes are sorted vertically by mean firing rate (sorted neuron ID).   Note that the resulting spike raster is reminiscent of the  letter `$\pi$'. The `left leg' comprises spikes shortly before (and thus associated with) the next NS.   The `right leg' comprises spikes long before the next NS and thus presumably associated with the last NS.  For all but the most active neurons, the two `legs' are distinct, so that `first spikes' can be identified without ambiguity (rightmost spikes in `left leg').

The first-spike latency of individual neurons, relative to the nearest NS, is illustrated in {\bf Figure~\ref{first_hist2} A}.   It is evident that firing latency decreases systematically with mean activity.  The least active neurons consistently fire {\it after} NS (positive latencies, ID $6$ to $55$).  Neurons with intermediate activity ($55<\mathrm{ID} < 260$) consistently fire {\it with} the NS (near-zero latencies).  (The ordering of these neurons is effectively random, as they exhibit identical levels of activity.)  More active neurons ($260 < \mathrm{ID} < 320$) fire consistently {\it before} the NS (negative latencies).  The most active neurons ($320 < \mathrm{ID}$) fire at all times (both positive and negative latencies).   `Pioneer neurons' are neurons with consistently negative latency.  A specific criterion foe `pioneers' is that the absolute value of mean latency be greater than the standard deviation of latency.


Why should firing latency grow more negative with higher mean activity?  A simple explanation is differential `sensitivity', that is, different probabilities that small fluctuations of population activity evoke a spike.  More `sensitive' neurons would be recruited more frequently and thus show higher mean activity.  By the same token, more `sensitive' neurons would be recruited earlier by the rising activity preceding a NS.  Accordingly, the observed link between firing latency and mean activity is consistent with differential `sensitivity'.

To test this hypothesis, we established in simulations the individual distribution of membrane potential $V$ for each excitatory neuron.   For convenience, we illustrate the results for a hypothetical potential $V^*$, which is identical to $V$ except in that it is never reset (and thus avoids discontinuities at threshold).  As expected, the distribution of $V^*$-potential shifted systematically with mean activity ({\bf Figure~\ref{first_hist2} B}).  In the range of `pioneers' ($260 < \mathrm{ID} < 320$), the mean value $\left\langle V^* \right\rangle$ was just about one standard deviation below threshold voltage $V_\mathrm{th}$.  Below this range, $\left\langle V^* \right\rangle$ was consistently and well below threshold and, above this range, $\left\langle V^* \right\rangle$ is consistently and well above threshold.   This confirms that `pioneers' are most `sensitive' (in the sense of being nearest to threshold) among excitatory neurons.

To further clarify the relation between individual neuron spikes and fluctuations of population activity, we computed the average population activity $\Gamma_i(\tau)$ preceding a single neuron spike, over and above mean population activity (see Methods, {\bf Figure~\ref{first_hist2} C}). To avoid contamination by NS, the analysis was based exclusively on periods  {\it between} NS ($1000 \,\, \mathrm{s}$ simulation).  The results revealed that spikes of the earliest `pioneers' ($280<\mathrm{ID}<320$) were preceded by positive deviations of population activity (amplitude $\sim 1.5 \,\, \mathrm{Hz}$, range $-60 \dots -40 \,\,\mathit{ms}$), or in other words, by $\sim 30$ additional excitatory spikes over a $\sim 20 \,\, \mathrm{ms}$ period.

Pioneers may be not just sensitive to, but also influential on, population activity.  To assess the potential influence of pioneers on population activity, we established the synaptic effectiveness over all efferent projections.  As a first step, we computed the probability density of recovered synaptic resources for the efferent projections of each excitatory neuron ({\bf Figure~\ref{depression} A}).   Unsurprisingly, synaptic resources proved to be more depleted in more active neurons.  However, pioneers, which rarely fire between NS, retained at least $75$ $\%$ of their synaptic resources.  As a second step, we assessed post-synaptic impact by computing the average post-synaptic potential elicited by a single spike of different neurons ({\bf Figure~\ref{depression} B}) and by repeated (Poisson) spiking of different neurons ({\bf Figure~\ref{depression} B}).   This analysis did not show `pioneer' neurons to be uniquely influential.  It did, however, show them to be the most active and sensitive neurons retaining substantial synaptic resources and therefore with influential single spikes.  In neurons that are more active, the influence of single spikes is smaller but the combined influence of all spikes is larger.

Following \citetext{Zbinden}, we tried to related the `influence' of individual pioneer spikes on subsequent population activity and the number of efferent projections of the neuron in question.  Although pioneers exhibitied widely different influence, we found no straightforward relation to connectivity ({\it e.g.}, in terms of a standard definition of `hubs').  Apparently, the `influence' of pioneers varies with effective connectivity, both direct and indirect, which is not easily quantified.


In a further attempt to assess `influentialness' , we selectively silenced different groups of excitatory neurons \cite{tsodyks2}. Remarkably, silencing early `pioneers' ($290 \leq \mathrm{ID} \leq 320$) eliminated large synchronization events ($>14.5$ times mean activity), leaving only far smaller activity fluctuations ($<5.0$ times mean activity) ({\bf Figure~\ref{de_eff_exp} A}).   Silencing either less active neurons ($\mathrm{ID} < 290$) or more active neurons  ($360 < \mathrm{ID}$) did not have this drastic effect.  Evidently, `pioneers' were uniquely influential in terms of initiating large synchronization events.  Interestingly, this cannot be attributed to disproportionate post-synaptic impact.  As mentioned, `pioneers' were not exceptional when post-synaptic impact of Poisson firing was compared ({\bf Figure~\ref{depression} B and C}).  Note, however, that this steady-state measure is unlikely to fully capture the `runaway' dynamics of NS initiation.


To assess the possibility of differential contributions to the dynamics of NS initiation, we estimated thresholds for NS initiation in partially silenced networks (see Methods).  To this end, we established the bimodal distribution of peak activities during fluctuations of spontaneous activity (cf. {\bf Figure~\ref{all-or-none} B}), which reveals a low range of smaller fluctuations and a high range of full-blown NS.  Threshold was defined as the largest observed value in the low range.   Silencing a group of neurons reduces effective connectivity and recurrent amplification, which is expected to elevate threshold for NS initiation.  If all neurons contribute similarly to amplification, silencing any group would elevate thresholds similarly.  If some neurons contribute more than others, silencing some groups would elevate threshold differentially.   The observed effect of silencing different groups of neurons is shown in  {\bf Figure~\ref{de_eff_exp} B}.  Clearly, the threshold of NS initiation was elevated disproportionately by silencing pioneer neurons.  Over much of the pioneer range, the threshold was elevated to `ceiling', in the sense that even the largest observed activity fluctuations failed to trigger a NS.



We conclude that `pioneers' are exceptional in combining high `sensitivity' to fluctuations of activity with large `influence' on such fluctuations.  High `sensitivity' is a consequence of the membrane potential hovering just below threshold and large `influence' is partially a consequence of largely recovered synaptic resources.  In combination, `sensitivity' and `influence' of individual `pioneer' spikes contribute disproportionately to the `runaway' dynamics of NS initiation, in that their silencing disproportionately elevates the threshold for NS initiation.  Importantly, both `sensitivity' and `influence' result from an intermediate level of mean activity, which is why `pioneers' comprise a contiguous range of activity-sorted neurons.    


\subsection{Order-based representation} 
\label{order_based}

This section compares the representation of external stimulation by different aspects of neuronal activity.  Following  \cite{kermany}, we consider two spike-based and two rate-based encoding schemes by a set of $n$ neurons.  The spike-based schemes are, firstly, the times $t_1, \ldots t_n$ of the first spike of each neuron following stimulation (`spike times') and, secondly, the rank-order $o_1, \ldots o_n$ of the first spike of each neuron following stimulation (`spike order').  The rate-based schemes are, thirdly, the mean spike rate $c_1, \ldots c_n$ of each neuron during the $100\,\,\mathrm{ms}$ following stimulation (`neuronal rates') and, fourthly, the temporal profile of the combined spike rate $r_1, \ldots, r_{50}$ of all $n$ neurons during $50$ successive time-bins of $2\,\,\mathrm{ms}$ duration following stimulation (`temporal rates').

External stimulation was delivered to $k$ groups of excitatory neurons, each with $s=10$ neurons.  Target neurons were chosen randomly.   We refer to each group of target neurons as a `stimulation site', although of course there is no spatial location in our simulated networks.   Stimulation was delivered at random sites and random times reflecting a Poisson process (cf. {\bf Figure~\ref{stimulation} A}).  Stimulation rate was set at $1 \,\, \mathrm{Hz}$, in order to obtain more evoked than spontaneous NS. Four heterogeneous random networks were simulated, under two conditions: $k=5$ sites and $300 \,\, \mathrm{s}$ simulation time, and $k=12$ sstimulation sites with $720 \,\, \mathrm{s}$ simulation time.  Only synaptically mediated spikes (not spikes enforced by stimulation) were included in the analysis.  Classification of stimulation sites proved to be non-trivial.

To assess quality of representation by different groups of neurons, we established classification performance separately for non-overlapping groups of $n$ activity-sorted neurons (sorted ID  $[1,\ldots,n]$,  $[n+1,\ldots,2n]$, $[2n+1,\ldots,3n]$, and so on).  For the easier task of decoding $k=5$ stimulation sites, we used sets of groups of $n=10$ neurons.  For the harder task of decoding $k=12$ stimulation sites, we used groups of $n=30$ neurons.  The observed classification performance is shown in {\bf Figure~\ref{resonance1}} A to D ($5$ stimulation sites) and {\bf Figure~\ref{resonance1}} E to H ($12$ stimulation sites).  Classification performance was far better for the two spike-based schemes (`spike time' and `spike order') than for the two rate-based schemes (`neuronal rates' and `temporal rates'). 
Interestingly, classification performance peaked when decoding was based (in part) on pioneer neurons.  This was more apparent for $k=5$ stimulation sites ({\bf Figure~\ref{resonance1}}, A to D), where performance peaked at intermediate values, than for $k=12$ stimulation sites ({\bf Figure~\ref{resonance1}}, E to H), where performance reached higher values (and thus suffered a ceiling effect).  Performance of the rate-based scheme barely exceeded chance.


The exceptional representational capacity of pioneer neurons became even more evident when the similarity of two `spike orders'  $o_1, \ldots o_n$ and  $o'_1, \ldots o'_n$ was quantified.  To this end, we modified the `Levenshtein edit distance' \cite{levenshtein} and measured `spike order similarity' (SOS)  by means of a measure based on permutations (see Methods).  Note that SOS may be computed for any pair of NS, spontaneous or evoked.   Sorting all observed NS by class (`spontaneous', `evoked at site 1', `evoked at site 2', and so on), the results were collected into the similarity matrices  shown in {\bf Figure~\ref{matrices} A, B, C, and D}.  To assess the representational capacity of pioneers, we computed such matrices both from the SOS of pioneers and non-pioneers. Pioneers were chosen randomly from the range $[260,320]$ ($k=5$ and $n=10$ in $\mathbf{C}$; $k=12$ and $n=30$ in $\mathbf{D}$). Non-pioneers were chosen randomly from the range $[0,260]$ ($k=5$ and $n=10$ in $\mathbf{A}$; $k=12$ and $n=30$ in $\mathbf{B}$). It is evident that SOS of non-pioneers was high between all NS ({\bf Figure~\ref{matrices} A and B}).  Apparently, non-pioneers were recruited in stereotypical order during all NS.   In contrast, the SOS of pioneers was generally high for NS evoked at the same site, but low for spontaneous NS and for NS evoked at different sites ({\bf Figure~\ref{matrices} C and D}).  Clearly, `pioneers' were exceptional in that their `spike order' was both unusually variable and unusually informative about stimulation site.

To corroborate these observations and to comprehensively compare all groups of excitatory neurons, we established the distribution of spike-order similarity (SOS), both within and between classes of NS.  As before, NS classes are understood to be `spontaneous', `evoked at site 1', `evoked at site 2', and so on. Given two distributions of SOS values, with means $\mu_{1,2}$ and standard deviations $\sigma_{1,2}$, we expressed their `distance' in terms of the z-score, $z ={|\mu_2 - \mu_1|}/({\sigma_1 + \sigma_2})$.  The results are shown in {\bf Figures \ref{matrices} E and \ref{matrices} F}, for the two experiments with $k=5$ and $k=12$ stimulation sites, respectively.  In both experiments, the difference between `within-class' and `between-class' distributions was most pronounced when SOS was computed for pioneers (ID 260 to 320).  This demonstrates conclusively that the `spike order' of pioneers was more informative about stimulation site than any other group of excitatory neurons.


\clearpage

\subsection{Broadly heterogeneous connectivity} 
 \label{broadly_heterogeneous}
 
Finally, we compare the role of connection topology in shaping the macroscopic and microscopic activity considered above. To ensure generality, we considered thousands of networks with different combinations of recurrent excitation and inhibition, such as to map out the dynamical characteristics in detailed `landscapes'.  To obtain comparable macroscopic and microscopic dynamics from different network architectures, while also keeping computational effort manageable, we developed a semi-analytic procedure for network configuration (see Methods \ref{semi-analytic}).

The reference point for this effort was the average connectivity $\omega_{\mathit{ee,0}}$, $\omega_{\mathit{ei,0}}$, $\omega_{\mathit{ie,0}}$, $\omega_{\mathit{ii,0}}$  of the heterogeneous random networks analyzed above.  Varying relative strength of excitation, $w_E$, and inhibition, $w_I$,
randomly in the range of $0.0$ and $2.0$, we evaluated the activity dynamics expressed by any given average connectivity, $\omega_\mathit{ee}=w_E \omega_{\mathit{ee,0}}$, $\omega_\mathit{ie}=w_E\omega_{\mathit{ie,0}}$ , $\omega_\mathit{ei}=w_I \omega_{\mathit{ei,0}}$, and $\omega_\mathit{ii}=w_I \omega_{\mathit{ii,0}}$, separately for homogeneous random, scale-free, and heterogeneous random networks.  Specifically, we established the mean inter-network-spike interval ($T_\mathrm{INSI}$), the coefficient of variation of the inter-network-spike interval ($\mathit{CV}_\mathrm{INSI}$), the ratio between maximal and mean activity ($A_\mathrm{max}/A_\mathrm{mean}$), and the fraction of excitatory neurons that were pioneers ($f_\mathrm{pioneer}$).

We obtained $T_\mathrm{INSI}$ and $\mathit{CV}_\mathrm{INSI}$ directly from detected activity peaks ($\theta = 0.5 A_{\mathit{max}}$, see Methods).  As activity fluctuations were not guaranteed to be distributed bimodally, $A_\mathit{max}/A_\mathit{mean}$ offered a convenient proxy for the presence of `all-or-none' synchronization events, as its value increases with the size of synchronization events and decreases with their frequency.  For the determination of $f_\mathrm{pioneer}$, pioneers were defined by the mean of first-spike latency being more negative than its standard deviation (see Methods \ref{semi-analytic}).

The resulting landscapes are shown in {\bf Figure~\ref{landscape}}.  All three network types are characterized by a transition (blue dashed curves) between dynamical regimes dominated by inhibition and excitation, respectively.  Above the transition, dominant inhibition created an `asynchronous' regime, in which synchronization events were small, frequent, and irregular (small $A_\mathit{max}/A_\mathit{mean}$, due to small $A_\mathit{max}$, small $T_\mathrm{INSI}$, large $\mathit{CV}_\mathrm{INSI}$).  Below the transition, dominant excitation created a `tonic' regime,  where synchronization events were large, frequent, and regular (small $A_\mathit{max}/A_\mathit{mean}$ again, but now due to large $A_\mathit{mean}$, small $T_\mathrm{INSI}$, small $\mathit{CV}_\mathrm{INSI}$).  

Of particular interest is the transition region, where synchronization events were large, infrequent, and  irregular  (large $A_\mathit{max}/A_\mathit{mean}$, large $T_\mathrm{INSI}$, large $\mathit{CV}_\mathrm{INSI}$).   This `all-or-none' regime most resembled the experimentally observed activity of {\it in vitro} networks \cite{eytan,shahaf,kermany}.  Importantly, pioneer neurons formed only in this transition region {\bf Figure~\ref{landscape})}.  

Although qualitatively similar `landscapes' were produced by all three network types, there was an important quantitative difference: heterogeneous random networks featured a broader transition region, with `all-or-none' synchronization events and pioneer neurons being expressed over a wider range of $w_E, w_I$ balances.  This is consistent with our observation that heterogeneous random networks expressed `all-or-none' dynamics more robustly and reliably than other networks, which tended to require fine-tuning of parameters.  We conclude that heterogeneous connectivity stabilizes both macroscopic and microscopic features of activity dynamics (`all-or-none' events and `pioneers').  



\clearpage 

\pagebreak

\section{Discussion}

Over the past decade, compelling evidence has come to light that even unstructured networks of cortical neurons {\it in vitro} are capable of encoding and propagating information about past external stimulation \cite{eytan,shahaf,kermany,Levy}.  Specifically, such networks appear to encode information in terms of the ordering of individual spikes from a privileged group of neurons, termed `pioneer neurons'.  If even unstructured, {\it in vitro} networks -- which, in contrast to structured cortical networks {\it in vivo}, have not been shaped by either neural development, sensory inputs, or reinforcement learning -- possess such representational capabilities, this may have considerable implications for our general understanding of neural function.  For cortical neuronal networks {\it in vivo} might well subserve their individual functions roles by exploiting, extending, or customizing such intrinsic representational capabilities.   Machine learning applications such as `reservoir computing' further underline the functional possibilities of unstructured, recurrent networks, at least in combination with suitable decoding schemes \cite{Maass2002,Jaeger2004,Lukosevicius2009}.

Here we show that the representational capabilities of unstructured networks observed {\it in vitro} emerge robustly under minimal assumptions.  Firstly, we show that a network of excitatory and inhibitory spiking neurons with frequency-dependent synapses robustly expresses `pioneer neurons', provided only that degree of connectivity varies broadly across the network.  Secondly, we show that `pioneer neurons' reliably represent the site of prior external stimulation, in that the order of individual spikes depends characteristically on stimulation site.   The same is not true for any other cohort of excitatory neurons.  In a forthcoming publication, we will report additionally that sparse projections from `upstream' to `downstream' networks can efficiently propagate the information encoded by `pioneer neurons' \cite{Poster2}.

Several previous studies have thematized the emergence of `pioneer neurons' in unstructured networks  \cite{tsodyks2,Vladimirski,Zbinden}.  In these studies, hetereogeneity among excitatory neurons was obtained by means of variable (effective) firing thresholds, whereas connectivity remained homogeneously random ({\it i.e.}, Erd\"os-Renyi type)\footnote{Note, however, that connection {\it number} still varies by approximately $10\%$ due to randomness.}.  Sorting excitatory neurons by average activity, Tsodyks and colleagues (\citeyear{tsodyks2}) showed that a cohesive cohort of neurons with intermediate activity fires reliably in advance of large synchronization events (`network spikes', NS).  The same authors suspected these `pioneer neurons' to operate close to firing threshold and showed that the deafferentiation of `pioneer neurons' reduces the rate of NS disproportionately.  Vladimirski and colleagues (\citeyear{Vladimirski}) analyzed the effect of heterogeneous firing thresholds with a mean-field description of collective dynamics, confirming the importance of neurons with intermediate activity and demonstrating that heterogeneity ensures comparable dynamical instability in different network realizations.  Finally, Zbinden (\citeyear{Zbinden}) sought to differentiate between neurons of intermediate activity in terms of afferent and efferent connectivity, reporting that influence on NS grows with the number of efferent projections.


We have studied the emergence of `pioneer neurons' in unstructured networks with heterogeneous connectivity.  Specifically, we compared networks with comparable connectivity for all neurons (`homogeneous random' or Erd\"os-Renyi), networks with a small fraction of highly connected neurons (`scale-free', \citeauthor{Barabasi}, \citeyear{Barabasi}), and networks with various degrees of connectivity (from zero to twice average) occurring equally often (`heterogeneous random').   Other than connectivity, no other sources of heterogeneity were introduced in our networks ({\it i.e.}, all excitatory neurons were modeled with identical firing thresholds and background currents).       

\begin{figure}[!ht]   \centering   \includegraphics[width=12cm]{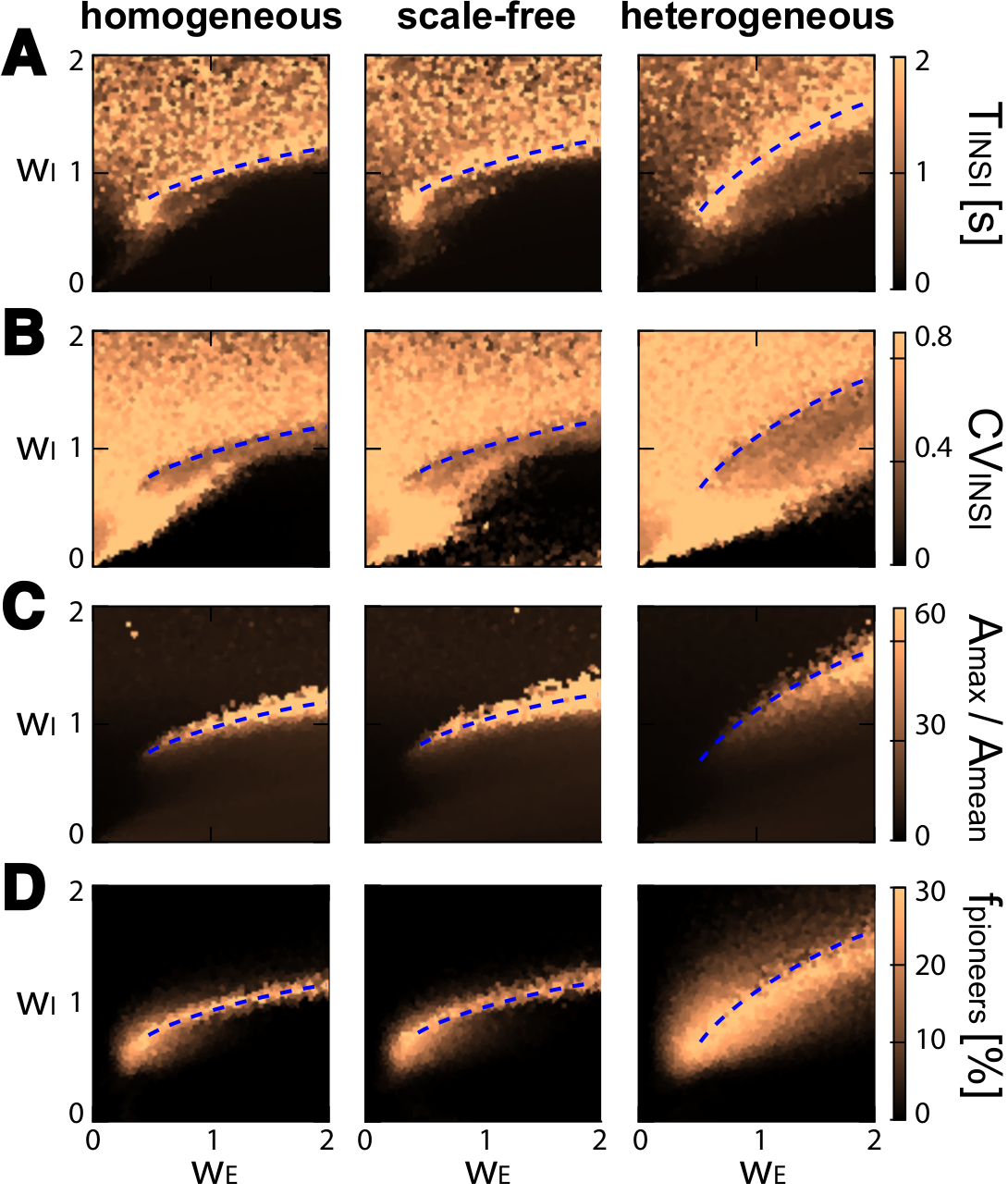}   \caption{Dynamical regimes of three network types,  as a function of excitation and inhibition. $\mathbf{A}$ mean NS interval $T_\mathrm{INSI}$ (in $\mathrm{s}$); $\mathbf{B}$ coefficient of variation of NS interval $\mathit{CV}_\mathrm{INSI}$; $\mathbf{C}$ activity ratio $A_\mathit{max}/A_\mathit{mean}$; $\mathbf{D}$ fraction of pioneers $f_\mathrm{pioneeer}$ (in $\%$).  Blue dashed curves indicate transitional region with `all-or-none' dynamics and numerous `pioneer neurons'.  This transitional region separates regime with `tonic' dynamics (below) from regime with `asynchronous' dynamics (above).  See text for further details.} \label{landscape} \end{figure}

All investigated network types expressed qualitatively similar macroscopic and microscopic dynamics.  All types generated network spikes (NS) at certain ratios of excitation and inhibition, as well as a cohort of moderately active neurons firing consistently prior to NS (Fig.~\ref{landscape}).   In quantitative terms, however, heterogeneous networks stood apart from the others: NS were expressed at lower levels, and over a wider range, of excitation (Fig.~\ref{landscape}A), the peak activity of NS was less extreme (Fig.~\ref{landscape}C), and `pioneer neurons' formed over a wider range of excitation and inhibition (Fig.~\ref{landscape}D).  In short, both macroscopic and microscopic phenomenology was expressed more consistently in heterogeneous networks, in the sense of being more reproducible both over different levels of excitation/inhibition and over different realizations of random connectivity. 

\begin{figure}[!ht]   \centering   \includegraphics[]{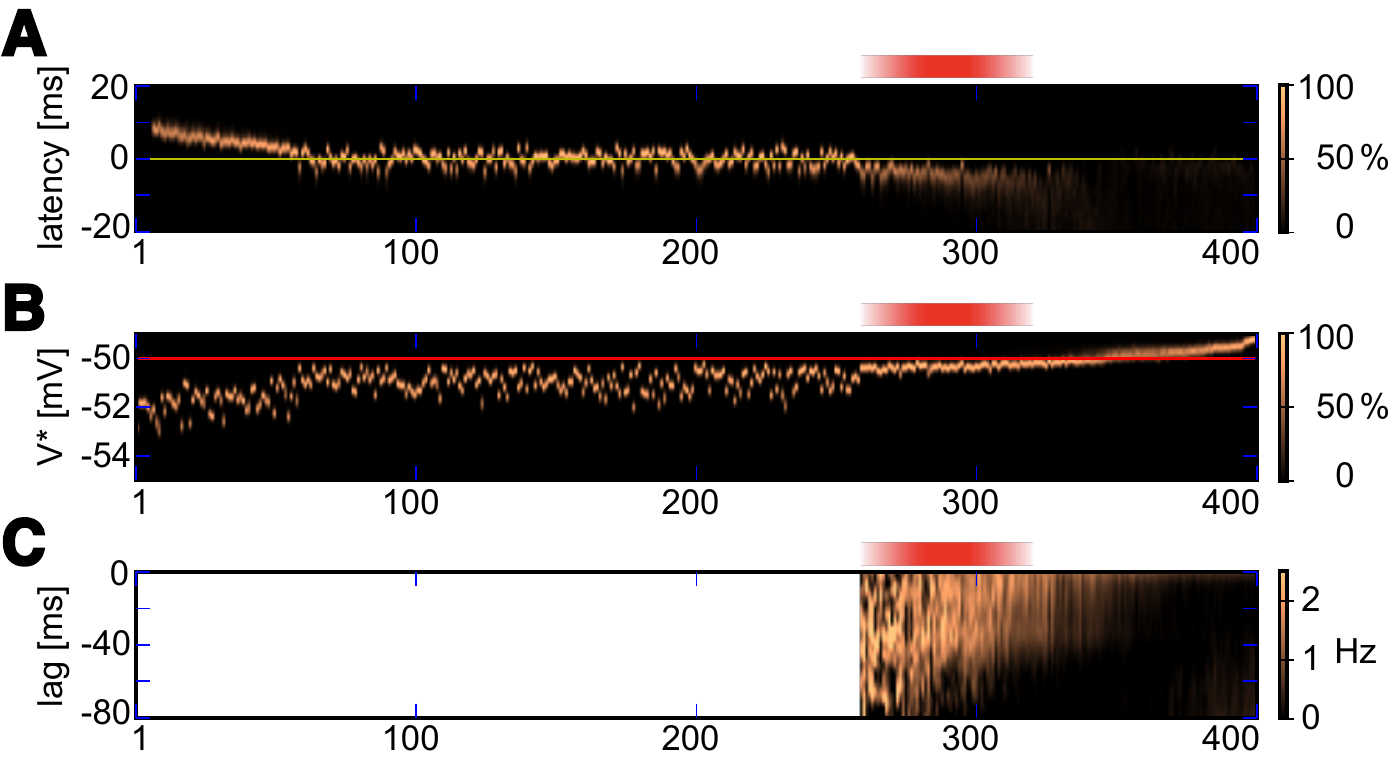}   \caption{Population activity, individual neuron spikes and membrane potential.  Excitatory neurons are sorted horizontally by mean activity (sorted neuron ID).  Red shading marks the pioneer range (ID 260 to 320).  $\mathbf{A}$: Latency of individual neuron first-spikes, relative to associated NS.  Zero latency (yellow line) is defined by peak activity of associated NS.  Neurons in the pioneer range fire reliably before the NS (negative latencies).  Color scale indicates fraction of maximal density.  $\mathbf{B}$: Distribution of $V^*$-voltage in individual neurons, relative to firing threshold (horizontal red line).  Neurons in the pioneer range have membrane potential just below threshold.  Color scale indicates fraction of maximal density.  $\mathbf{C}$: Average deviation $\Gamma_\mathrm{i}(\tau)$ of population activity at lag $\tau$, conditioned on individual spikes of neuron $i$ (see text and Methods).  Spikes of neurons in the pioneer range are consistently preceded by positive deviations.  Note that deviation $\Gamma_\mathrm{i}(\tau)$ is not defined below ID $260$.} \label{first_hist2} \end{figure}

We investigated in detail the distinguishing features of `pioneer neurons'  and how they relate network connectivity and dynamics.  By definition, `pioneers' differ from other excitatory neurons in that their spikes consistently {\it precede} NS (Fig.~\ref{first_hist2}A).   Another difference is that `pioneer' spikes consistently {\it follow} positive fluctuations of population activity (Fig.~\ref{first_hist2}C). This unique sensitivity is due to the fact that the membrane potential of `pioneers' hovers just below firing threshold (Fig.~\ref{first_hist2}B).  This potential regime is due, in turn, to an intermediate degree of `effective afference'.   Provided network connectivity is sufficiently heterogeneous, a sizeable cohort of neurons is bound to possess the requisite degree of `effective afference'.  If this is not the case (such as in homogeneous random or scale-free networks), far fewer neurons are found in the regime just below threshold.

\begin{figure}[!ht]   \centering   \includegraphics[]{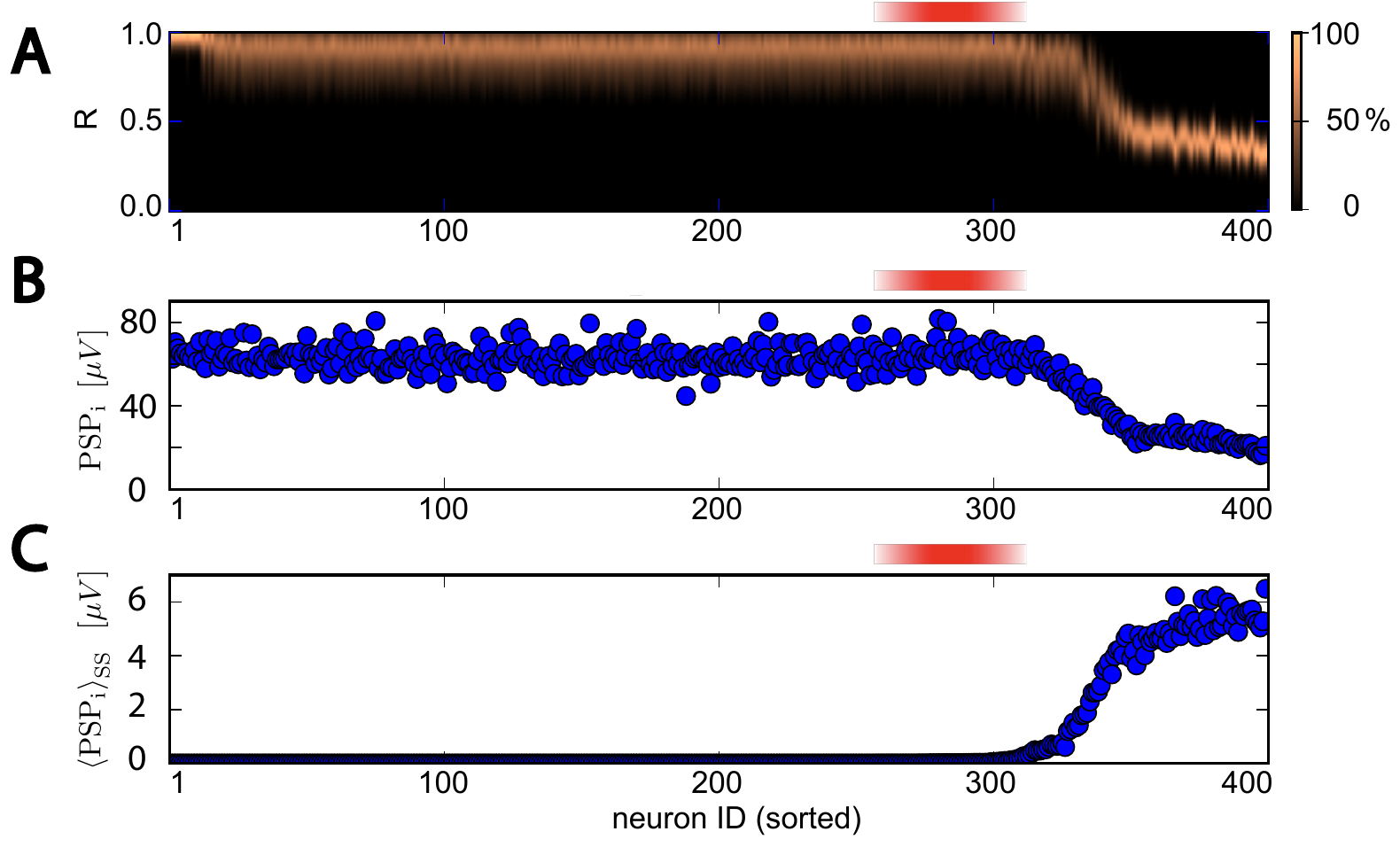}   \caption{Post-synaptic influence of excitatory neurons.   Red shading marks the pioneer range (ID 260 to 320).  $\mathbf{A}$: Probability density of synaptic resources $R$, average over all efferent synapses of a given neuron.  Resources decrease monotonically with mean activity.  The most active neurons to retain substantial resouces are neurons in the pioneer range.  $\mathbf{B}$  Amplitude of post-synaptic potentials $\mathrm{PSP}_\mathrm{i}$ elicited by single spikes, averaged over all efferent synapses.   $\mathbf{C}$: Steady-state post-synaptic potential $\left\langle\mathrm{PSP}_\mathrm{i}\right\rangle_\mathrm{ss}$ elicited by Poisson spiking at individual mean rate of neuron, averaged over all efferent synapses.  Note that increasing firing rate overcompensates diminishing resources.} \label{depression} \end{figure}

\begin{figure}[!ht]   \centering   \includegraphics[]{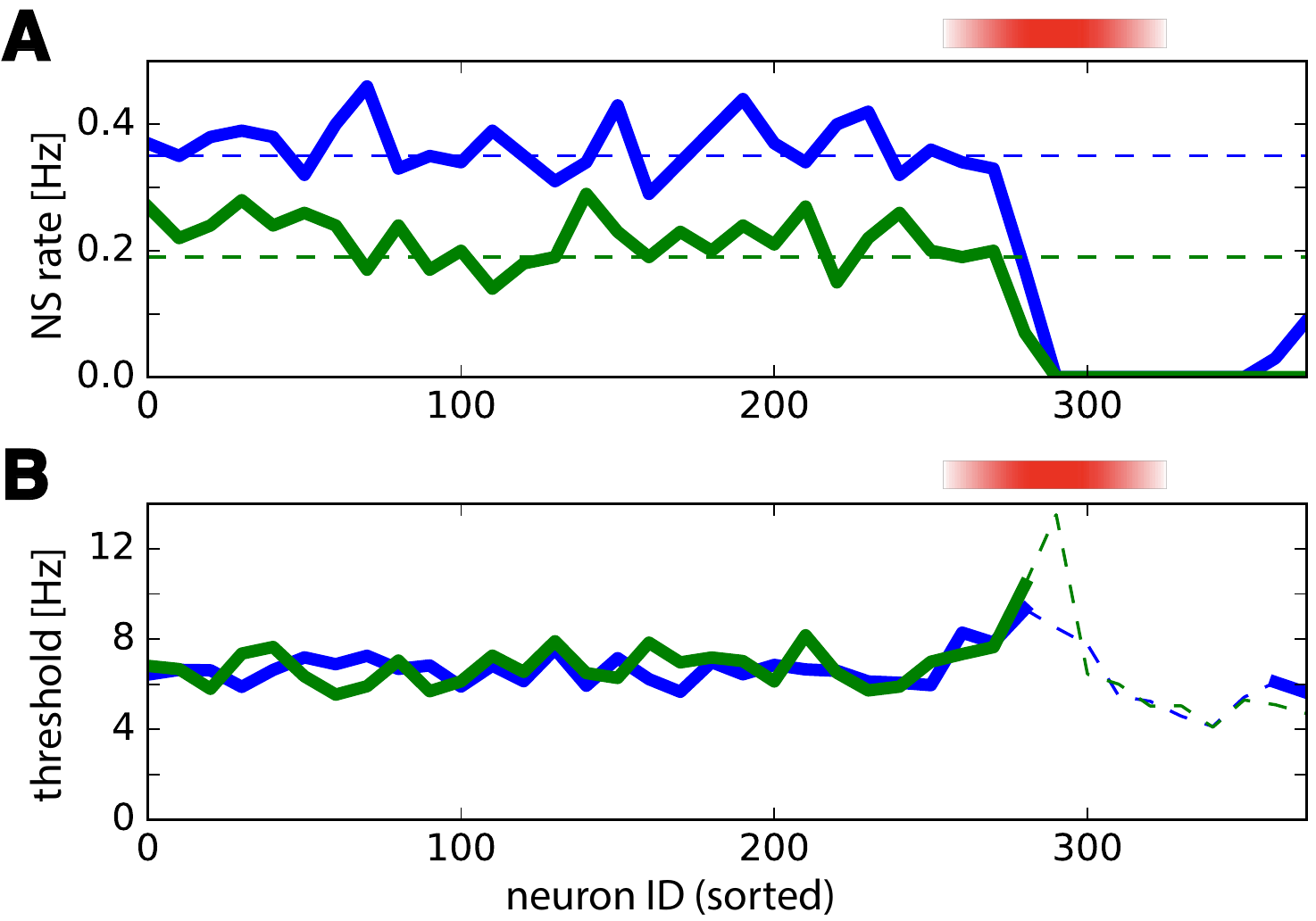}   \caption{Silencing pioneer neurons eliminates NS. Red shading marks the pioneer range (ID 260 to 320).  $\mathbf{A}$:  Rate of NS as a function of $N$, for modified networks with neuron cohort $N \leq \mathrm{ID} \leq N+30$ silenced by de-efferentiation.  Results are shown for two realizations (blue and green) with different spontanous NS rates (dashed lines).   NS cease when neurons in the pioneer range are silenced.  Typically, NS are recovered when neurons above this range are silenced ({\it e.g.}, blue trace). $\mathbf{B}$: Silencing pioneer neurons elevates threshold for triggering NS.  Threshold population activity (in $\mathrm{Hz}$), after silencing neurons $N \leq \mathrm{ID} \leq N+30$.  Two realizations are shown (blue and green traces).   Over much of the pioneer range, only a lower bound for the threshold could be established (dashed traces), because even the largest observed fluctuations failed to trigger a NS.} \label{de_eff_exp} \end{figure}

Note that we speak of {\it effective} afference or efference, because these do not reflect direct projections ({\it e.g.}, number and/or strength) so much as the non-local neighborhood of connectivity and activity.  For example, whereas average activity (used for sorting) is by definition a function of `effective' afference, it correlates neither with the number/strength of direct afferent inputs, nor even with the sum of the products of afferent strength and average pre-synaptic activity.   We suspect that a mean-field analysis of the non-local network neighborhood would be required to predict `effective' afference or efference.  Thus, in contrast to other network models \cite{Zbinden,Effenberger2015}, our `pioneers' are {\it not} distinguished by an unusually high degree of direct efferent connectivity, 

The importance of `pioneers' to NS becomes clear when connectivity is modified such as to selectively de-efferentiate different cohorts of excitatory neurons.   When the contribution of `pioneers' is removed in this way, the rate of spontaneous NS decreases, and the threshold for NS initiation increases, disproportionately (Fig.~\ref{de_eff_exp}).  The reasons are not immediately apparent.  Indeed, the afferent projections of `pioneers' retain most of their synaptic resources, due to the moderate activity of `pioneers' (Fig.~\ref{depression}A).  Thus, individual spikes of `pioneers' are more influential than spikes of other neurons with higher activity and more depleted synaptic resources (Fig.~\ref{depression}B).  However, averaging over the multiple spikes emitted during a NS, the impact of `pioneers' is smaller than that of more active neurons (Fig.~\ref{depression}B).

We conclude that the crucial role of `pioneers' for collective dynamics is due to a unique combination of sensitivity and influence.  Specifically, during the initiation phase of NS, individual `pioneer' spikes form the first links of the positive feedback chain sustaining the runaway dynamics of global synchronization.   This conclusion is consistent with all of our observations, including elevation of NS thresholds after `pioneer' de-efferentiation.   Moreover, it agrees with the analysis of Vladimirski and colleagues (\citeyear{Vladimirski}) that dynamical bistability is underpinned by ``a critical subpopulation of intermediate excitability conveys synaptic drive from active to silent cells''.  

\begin{figure}[!ht]   \centering   \includegraphics[]{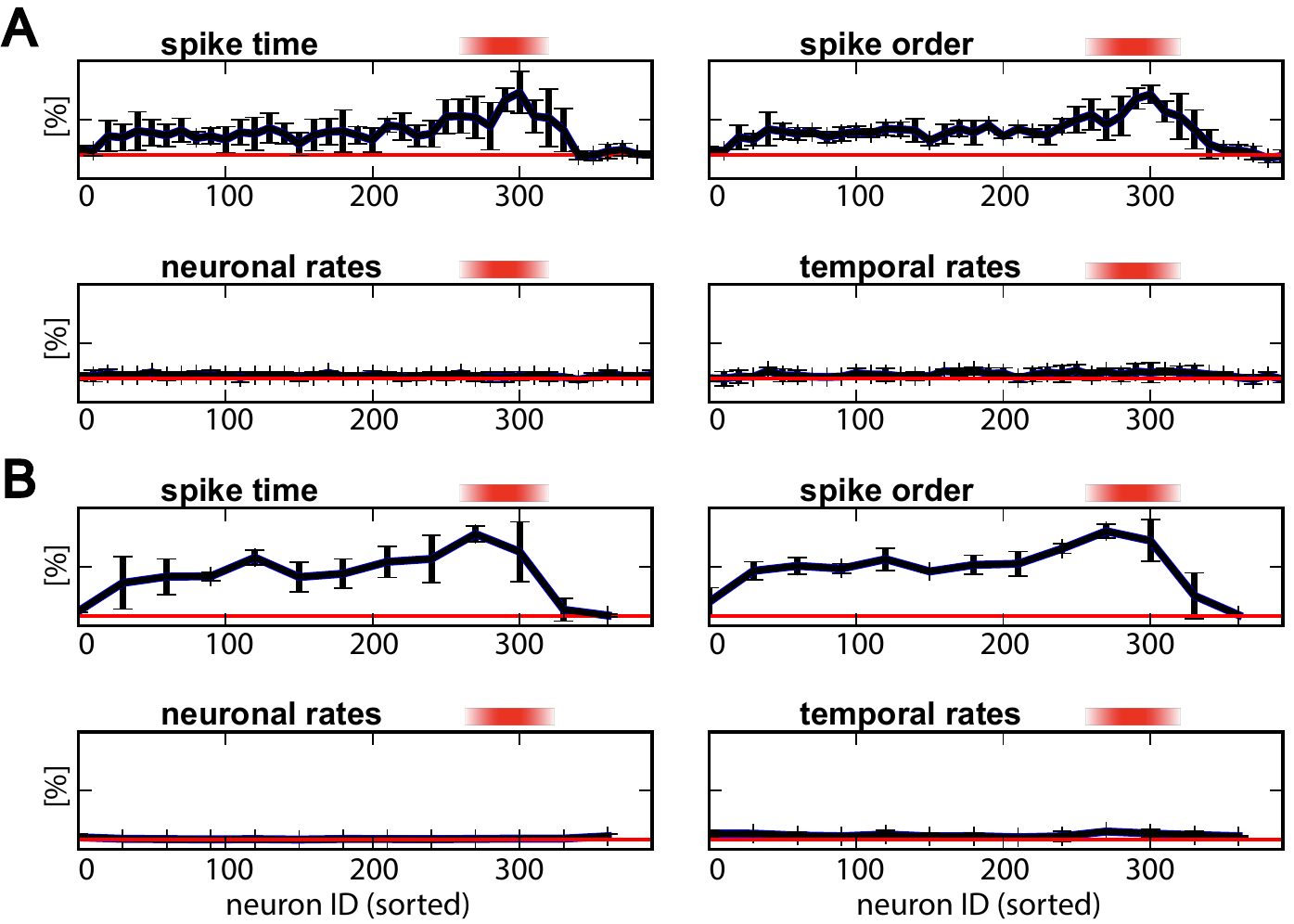}   \caption{Classification performance of different decoding schemes, based on different groups of neurons.  Red shading marks the pioneer range (ID 260 to 320).  Results for `spike time', `spike order', `neuronal rates', `temporal rates' are shown separately.  $\mathbf{A}$ Percentage of correct classification $\alpha(N)$ of one of five stimulated locations, based on the activity of neurons with sorted ID $[N,N+10]$.  Chance performance is 20\%.  $\mathbf{B}$  Percentage of correct classification $\alpha(N)$ of one of twelve stimulated locations, based on the activity of neurons with sorted ID $[N,N+30]$.  Chance performance is 8\%.} \label{resonance1} \end{figure}
\begin{figure}[!ht]   \centering   \includegraphics[width=11.0cm]{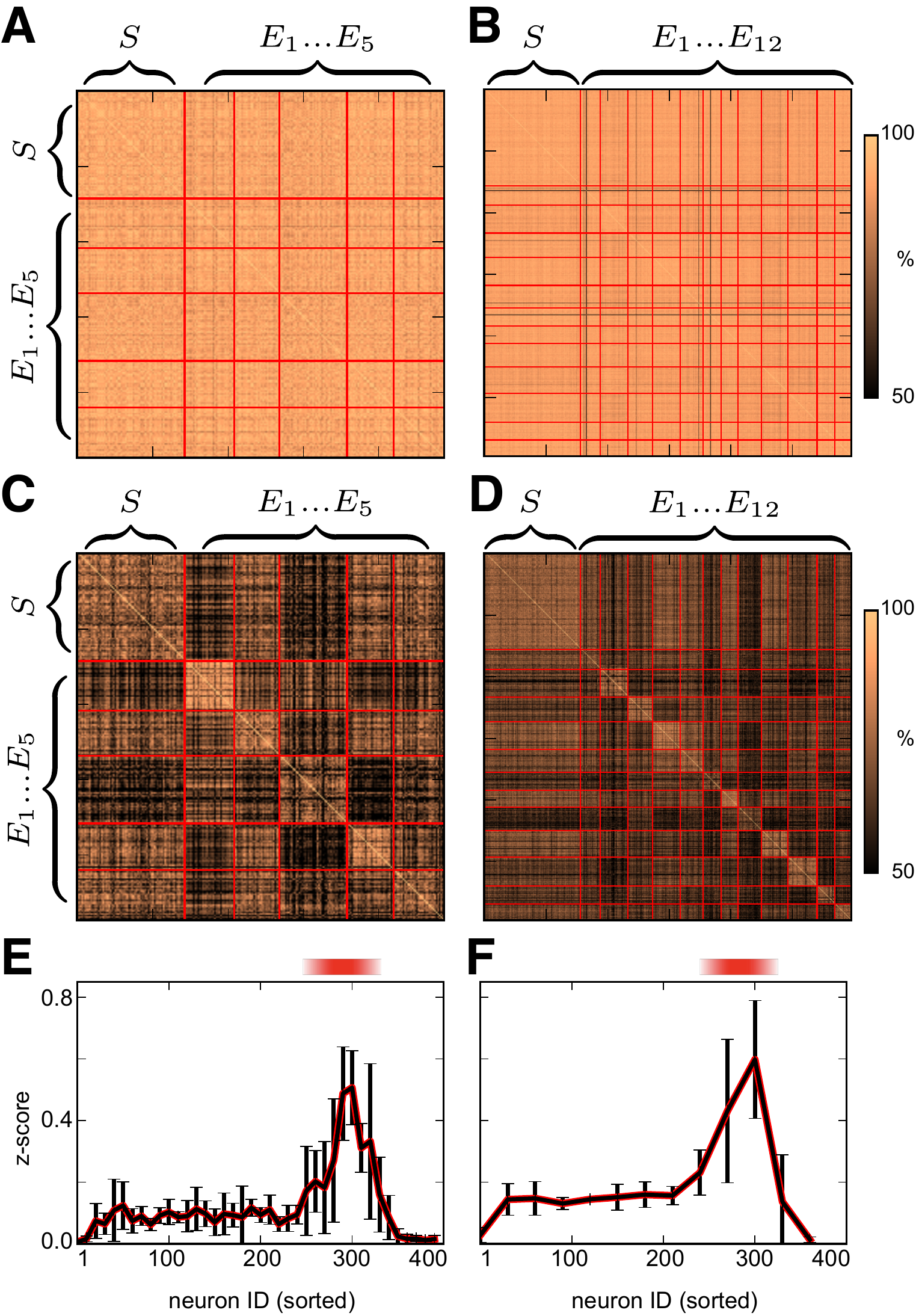}   \caption{Matrices of `spike order similarity' (SOS) during different NS.  Observed NS were classified as `spontanous' (S), `evoked at site 1' (E$_1$), `evoked at site 2' (E$_2$), and so on, and class boundaries between sorted NS are marked bu red lines.  $\mathbf{A-D}$ illustrate results for representative sets of non-pioneer or pioneer neurons. Color scales indicate fraction of maximal similarity.   $\mathbf{EF}$ illustrate results for all sets of neurons with contiguous ID.   $\mathbf{A}$: Non-pioneer SOS, five stimulation sites. $\mathbf{B}$:  Non-pioneer SOS, twelve stimulation sites.  $\mathbf{C}$: Pioneer SOS, five stimulation sites. $\mathbf{D}$:  Pioneer SOS, twelve stimulation sites.  $\mathbf{E}$:  Distance between SOS distributions, within-class and between-class, for five stimulation sites and sets of ten neurons (contiguous ID in range $[N,N+9]$).  $\mathbf{F}$:  Distance between SOS distributions, within-class and between-class, for twelve stimulation sites and sets of thirty neurons (contiguous ID in range $[N,N+29]$).} \label{matrices} \end{figure}

Finally, we investigated the representational capacity of `pioneers neurons', prompted by experimental evidence to this effect from unstructured networks {\it in vitro} \cite{kermany}.   Importantly, we found the ordering of `pioneer' spikes during NS initiation to be highly context-dependent: ordering was largely random before spontaneous NS, but turned more stereotypical before NS evoked by external stimulation (Fig.~\ref{matrices}CD).  Moreover, stereotypical spike ordering was characteristic for each of several alternative stimulation sites, reliably identifying the stimulation site (Fig.~\ref{resonance1}).  Other measures of population activity, such as the average activity profile of individual neurons (`neuronal rates') or the temporal profile of population activity (`temporal rates') were largely uninformative.

Surprisingly, spike ordering during NS initiation revealed a clear-cut difference between `pioneer' and `non-pioneer' neurons.  Apart from `pioneers', no other cohort of excitatory neurons showed any context-dependence in their spike order.  Instead, the recruitment order of other neuron cohorts remained highly stereotypical during NS initiation, whether `spontaneous' or `evoked' (Fig.~\ref{matrices}AB, Fig.~\ref{resonance1}).  We conclude that the representational capacity of `pioneer neurons' reflects a unique dynamical feature, namely, context-dependent recruitment order.

\begin{figure}[!ht]   \centering   \includegraphics[width=12cm]{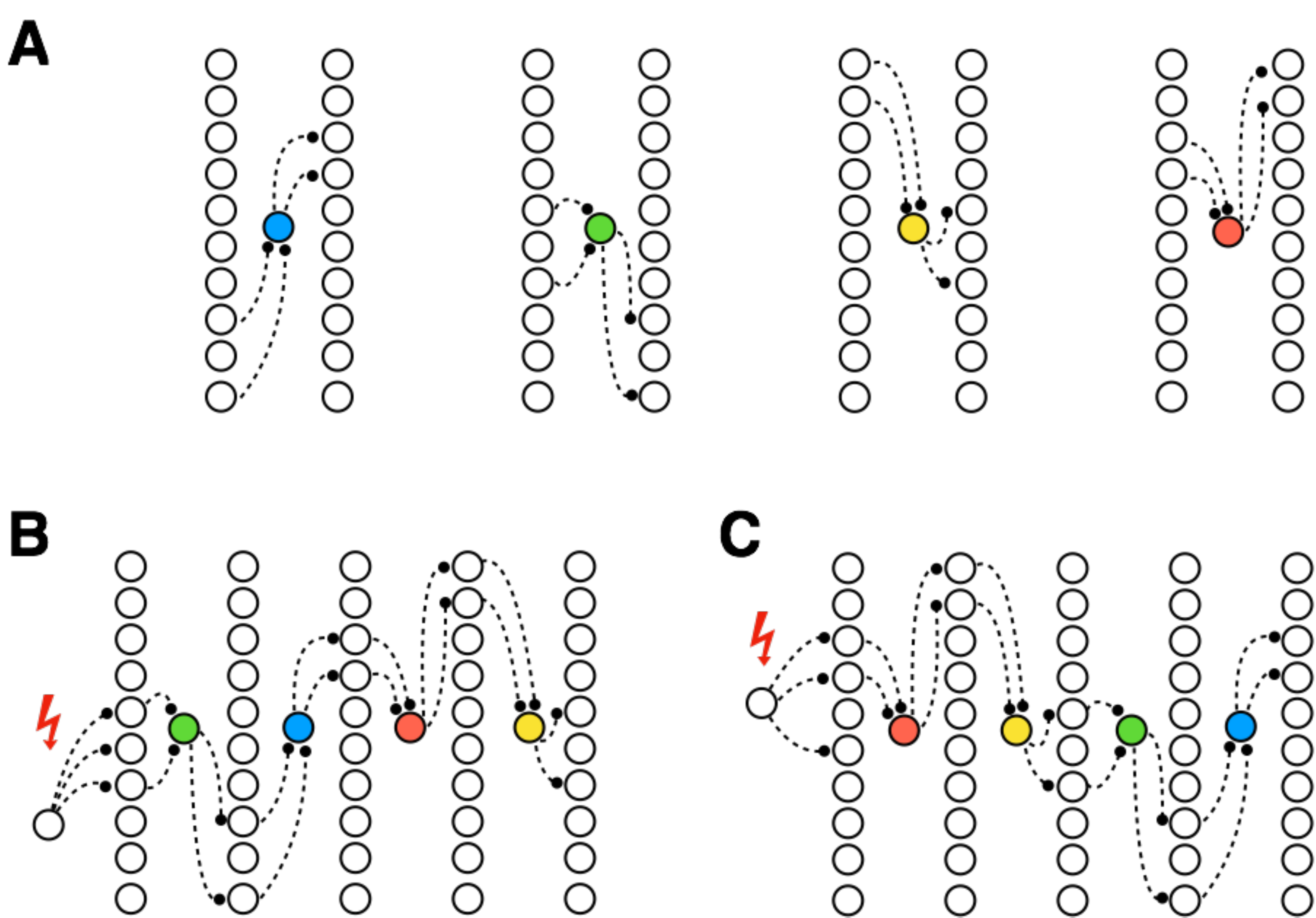}   \caption{Many-to-one-to-many propagation of activity by pioneer neurons (highly schematic). $\mathbf{A}$ Four pioneer neurons are illustrated (blue, green, yellow, red), receiving afferent input from the left, and emitting efferent output to the right.  Vertical columns of neurons represent the network as a whole.  Afferent and efferent projections involve independent and random subpopulations of the network.   $\mathbf{B}$ External stimulation (lightning) of a specific subpopulation propagates, via a particular `pioneer', to another subpopulation, starting an orderly sequence (green$\to$blue$\to$red$\to$yellow).  $\mathbf{C}$ External stimulation of another subpopulation propagates via another pioneer, starting another orderly sequence (red$\to$yellow$\to$green$\to$blue).} \label{chain_reaction} \end{figure}

We suspect that this dynamical feature results directly from the combined sensitivity and influence of `pioneers'.  To see the possible connection, it is important to realize that many spikes are needed to elicit a `pioneer' spike\footnote{{\it e.g.}, approximately $30$ additional population spikes precede each pioneer spike, Fig.~\ref{first_hist2}C} which, in turn, may elicit many subsequent spikes (``many-to-one-to-many'').  As both afferent and efferent connectivity is partial and random, a `pioneer' may be sensitive to one subpopulation and may convey activity to an independent subpopulation (Fig.~\ref{chain_reaction}A).  This `many-to-one-to-many' propagation would be effective only in a short window of time, before population activity has risen too high.  During this window, `pioneer' spikes would recruit subpopulations, which would in turn recruit other `pioneer' spikes, and so on, in a largely orderly sequence prescribed by afferent and efferent connectivity (Fig.~\ref{chain_reaction}BC).  In the absence of external stimulation, random fluctuations would determine the initiation point of this orderly sequence, scrambling the recruitment order.  External stimulation would prescribe a specific initiation point, resulting in a more reproducible sequence that would be informative about the stimulation site.


But how to explain that `non-pioneer' neurons are recruited in a fixed order?  The majority of excitatory neurons (ID 80 to 260) are less sensitive than `pioneers' ({\it i.e.}, with membrane potentials further from threshold) and are recruited only when a NS is well underway and population activity is already higher (Fig.~\ref{first_hist2}AB).  At this point, any information about the initial growth of activity is likely to have been washed out by surging excitation.  This would explain why the majority of `non-pioneer' neurons spike with largely deterministic latencies (positive or negative) relative to the peak of the NS and thus in largely deterministic order.  It would also explain why the majority of excitatory neurons carry little or no information about the site of external stimulation.




In conclusion, we have presented a minimal model for the emergence of a privileged class of neurons in unstructured networks, which provides a highly efficient order-based representation of external inputs.  We term this model `minimal' because it assumes only broadly heterogeneous connectivity in addition to standard neuron and synapse models (integrate-and-fire-neurons and frequency-dependent conductance synapses).  Thus, our findings do not depend in any way specific features such customized connectivity, inhomogeneous or dynamic background currents, or inhomogeneous or dynamic firing thresholds \cite{Persi2004a,Gritsun2011,Masquelier,gigante,Rajan2016}.


We believe that our findings offer a deeper understanding of both the mechanisms underlying, and the possible functional significance of, repeating `motifs' in the sequence of neuronal recruitment, as experimentally observed both {\it in vitro} \cite{eytan,Rolston2007,shahaf,kermany} and {\it in vivo} in sensory cortex \cite{Luczak2007,Luczak2012a,Luczak2012b}, prefrontal and parietal cortex \cite{Peyrache2010,Rajan2016}, and in hippocampus \cite{Matsumoto2013,Stark2015}.
We show how such repeating 'motifs' can result from a local interaction of cellular and synaptic conductances, as hypothesizes by several authors \cite{Luczak2012b,Rajan2016}, and demonstrate their potential functional significance as a highly compact and efficient representation of previous external inputs \cite{Contreras2013,Stark2015,Rajan2016}.

\pagebreak

\newpage  
\bibliographystyle{jneurosci}
\bibliography{paper}{}

\newpage

\end{document}